\documentclass[a4paper,aps,prd,10pt,preprintnumbers,showpacs,showkeys,
twocolumn,superscriptaddress,nofootinbib,amsmath,amssymb]{revtex4-1}
\usepackage{cmap}
\usepackage[utf8]{inputenc}
\usepackage[T1]{fontenc}
\usepackage{graphicx,orcidlink,xcolor}
\usepackage{soul}
\def\Order#1{{\cal O}\left(#1\right)}

\begin{document}

\title{Primary Proca Hair and the Double-Peak Optics of Black Holes}

\author{R. A. Konoplya \orcidlink{0000-0003-1343-9584}}
\email{roman.konoplya@gmail.com}
\author{D. Ovchinnikov \orcidlink{0000-0002-7811-7554}}
\email{dmitriy.ovchinnikov@physics.slu.cz}
\author{J. Schee}
\email{jan.schee@physics.slu.cz}

\affiliation{Research Centre for Theoretical Physics and Astrophysics, Institute of Physics, Silesian University in Opava, Bezručovo náměstí 13, CZ-74601 Opava, Czech Republic}

\begin{abstract}
We study the optical properties of black holes endowed with primary Proca hair, focusing on the distinctive double-peak structure generated in the effective potential by the massive vector field. This feature drastically modifies the geodesic motion of both photons and massive particles, leading to qualitatively new dynamical and observational signatures. We derive and analyze the effective potentials, classify time-like and null geodesics, and identify the conditions for multiple circular orbits. Particular attention is devoted to the photon sphere structure, the associated shadows, and lensing phenomena. Our analysis reveals that, for a broad range of parameters, the black-hole shadow can acquire a two-boundary structure and exhibit additional inner rings, unlike the standard Schwarzschild case. These modifications provide potentially observable imprints of Proca hair in electromagnetic spectra, highlighting the relevance of double-barrier optical phenomena for current and future observations of strong-gravity environments.
\end{abstract}


\maketitle

\section{Introduction}

Theories with higher-curvature corrections, such as Gauss–Bonnet gravity and its extensions, attract considerable attention as they naturally arise in the low-energy limit of string theory and offer a framework for probing quantum aspects of gravity beyond general relativity. Within these models, black-hole solutions often exhibit nontrivial features absent in Einstein theory, including the appearance of additional parameters that characterize the geometry independently of mass, charge, or angular momentum. Such \emph{primary hairs} provide a valuable testing ground for exploring how fundamental corrections to the gravitational action can alter the structure of horizons, the dynamics of perturbations, and the associated observational signatures. In this context, studying optical phenomena around hairy black holes offers an opportunity to connect higher-curvature gravity with potentially observable effects in astrophysics.  

An interesting model that admits new primary hairs was recently proposed by Charmousis et al.~\cite{Charmousis:2025jpx}. It incorporates a second-order curvature term of the Gauss–Bonnet type together with a Proca field that carries the primary hair. As shown in \cite{Konoplya:2025uiq}, black hole solutions in this theory exhibit qualitatively new features in their perturbation spectra, suggesting that the presence of primary hairs may lead to strong echoes. In this work, we are interested in exploring how such black holes could manifest themselves in electromagnetic observations.

Recent observational breakthroughs in the electromagnetic spectrum have opened a direct window onto the near-horizon structure of black holes. The imaging of black-hole shadows by the Event Horizon Telescope \cite{EventHorizonTelescope:2019dse,EventHorizonTelescope:2022wkp,EventHorizonTelescope:2019ggy,EventHorizonTelescope:2019pgp}, along with complementary studies of gravitational lensing, photon rings, and time-delay effects \cite{Cunha:2018acu,Perlick:2021aok}, has provided unprecedented opportunities to confront theoretical models with data. Shadows in particular encode essential information about the geometry and possible deviations from general relativity, making them a key probe of both classical and quantum-corrected black-hole spacetimes. Recently observed polarization flips also suggest the presence of time-dependent vector fields in the vicinity of a black hole~\cite{Akiyama:2025aa55855}.

The key optical and dynamical properties of black holes are determined by the radii of stable and unstable circular orbits for photons and massive particles, the size of the shadow cast by the black hole, the associated time delays, and the deflection angles of light. While some of these aspects (the shadow radius and the position of innermost stable circular orbit) were formally addressed in \cite{Lutfuoglu:2025ldc}, the restricted parameter range studied there overlooked a number of qualitatively new phenomena arising from the existence of a double-peaked effective potential, in which the secondary peak may even exceed the primary one known from Einstein gravity.

Here we demonstrate that a detailed analysis of basic optical phenomena — such as gravitational lensing, time delays, and shadows — reveals qualitatively different behaviors across certain parameter ranges, thereby suggesting rather stringent constraints on the parameters of the new theory.

The paper is organized as follows. In Sec.~II we introduce the background geometry with Proca hair and establish the theoretical framework. In Sec.~III we derive the equations of motion for time-like and null geodesics. Sec.~IV is devoted to a systematic analysis of geodesic motion, including stability conditions and the appearance of critical orbits. In Sec.~V we explore the consequences for optical phenomena, focusing on time delays, lensing, swept angles, and shadow properties. Finally, Sec.~VI summarizes our findings and discusses their broader implications.

\section{Background geometry and theoretical framework}\label{sec:background}

A broad class of spherically symmetric black-hole solutions was recently derived in~\cite{Charmousis:2025jpx}, within a setup that blends scalar–tensor and vector–tensor interactions of the Gauss–Bonnet type. The starting point is the action
\begin{equation}\label{action}
S = \int d^4x \sqrt{-g} \left( R - \alpha\, \mathcal{L}^{\text{VT}}_G - \beta\, \mathcal{L}^{\text{ST}}_G \right),
\end{equation}
where $\alpha$ and $\beta$ are two independent couplings. The relevant Lagrangians are
\begin{equation}
\mathcal{L}_{\mathcal G}^{\rm VT} = 4 G^{\mu \nu}W_\mu W_\nu + 8 (W_\rho W^\rho)\, \nabla_\mu W^\mu + 6(W_\rho W^\rho)^2,
\end{equation}
for the vector–tensor interaction of a Proca field $W_\mu$, and
\begin{equation}
\mathcal{L}_{\mathcal G}^{\rm ST} = \phi \mathcal{G} - 4 G^{\mu \nu} \nabla_\mu \phi \nabla_\nu \phi - 4 \Box \phi \, (\partial \phi)^2 - 2 (\partial \phi)^4,
\end{equation}
for the scalar–tensor counterpart with scalar $\phi$~\cite{Fernandes:2021dsb}. Here
\begin{equation}
\mathcal{G} = R^2 - 4 R_{\mu\nu}R^{\mu\nu} + R_{\mu\nu\alpha\beta}R^{\mu\nu\alpha\beta},
\end{equation}
is the Gauss–Bonnet term, while the Einstein tensor is given by
\begin{equation}
G_{\mu\nu} = R_{\mu\nu} - \tfrac{1}{2} g_{\mu\nu}R.
\end{equation}

A distinctive property of this construction is the appearance of solutions that are both asymptotically flat and regular, with the geometry being supported by the Proca vector independently of the ADM mass. The black-hole line element takes the standard spherically symmetric form
\begin{equation}\label{eq:line_element}
ds^2 = - f(r)\, dt^2 + \frac{dr^2}{f(r)} + r^2 \left(d\theta^2 + \sin^2\theta\, d\phi^2\right),
\end{equation}
with the specific form of $f(r)$ depending on the relation between $\alpha$ and $\beta$.  

When the couplings are unequal, $\alpha \neq \beta$, the solution has a square-root structure,
\begin{align}\label{metricfunc}
f(r) &= 1 - \frac{2 \alpha (M - Q)}{r(\alpha + \beta)} + \frac{r^2}{2(\alpha + \beta)} \\\nonumber
&\quad - \frac{r^2}{2(\alpha + \beta)}
\sqrt{1 + \frac{8 \alpha Q}{r^3} + \frac{8 \beta M}{r^3} - \frac{16 \alpha\beta (M - Q)^2}{r^6}}\,,
\end{align}
where $M$ is the ADM mass, and $Q$ is an additional integration constant tied to the Proca field. Importantly, $Q$ is not fixed by $M$ and therefore represents \emph{primary hair}: an intrinsic attribute of the black hole which cannot be reduced to a conventional conserved charge. The special case $Q=M$ recovers the black-hole solutions of the scalar–tensor sector in four-dimensional Einstein–Gauss–Bonnet gravity~\cite{Lu:2020iav,Kobayashi:2020wqy,Fernandes:2020nbq}.  

In contrast, when $\beta = -\alpha$, the square root disappears and the metric simplifies to a rational function,
\begin{align}
\lim_{\beta \to -\alpha} f(r) &= \frac{r^3}{r^3 - 4\alpha(M-Q)}\Bigg(1 - \frac{2M}{r}
\\\nonumber
&\quad + \frac{4\alpha (M-Q)^2}{r^4} - \frac{4\alpha(M-Q)}{r^3}\Bigg).
\end{align}
Regularity requires that the denominator remain finite outside the horizon. Both branches reduce smoothly to the Schwarzschild solution when $\alpha = \beta = 0$.

The horizon radius $r_h$ is obtained from the condition $f(r_h)=0$, and the existence or absence of horizons depends sensitively on the parameter set $(\alpha,\beta,M,Q)$. At large distances the metric function behaves as
\[
f(r) = 1 - \frac{2M}{r} + {\Order r^{-2}},
\]
verifying that $M$ indeed corresponds to the ADM mass~\cite{Arnowitt:1960es}.  

The parameter space of admissible solutions has been explored in detail in~\cite{Charmousis:2025jpx}. It was shown there that sufficiently large and positive couplings $\alpha,\beta$ can obstruct horizon formation in the generic case $\alpha+\beta \neq 0$, while in the rational branch $\alpha+\beta=0$, overly large values of the Proca charge $Q$ can have a similar effect. In these situations the solutions no longer describe black holes, but instead correspond to naked singularities or horizonless compact configurations, depending on the parameters.

The most striking property of this metric, first pointed out in \cite{Konoplya:2025uiq}, is its non-monotonic behavior within certain regions of the parameter space. This feature gives rise to a second peak in the effective potential for test fields propagating in the black-hole background and, by analogy, should also generate a similar additional peak in the effective potential governing particle motion. Although geodesics in this spacetime have been analyzed in \cite{Lutfuoglu:2025ldc}, this particular effect was not observed. The emergence of a second peak implies the existence of an additional photon orbit and leads to qualitatively new optical phenomena. This provides strong motivation for a detailed investigation of optical properties of black holes with primary hair.

\section{Equations of motion}\label{sec:eq_motion}
We consider a static, spherically symmetric metric~\eqref{eq:line_element}. Due to spherical symmetry the motion can be restricted to the equatorial plane $\theta=\pi/2$. The existence of the Killing vectors $\partial_t$ and $\partial_\phi$ implies the conserved quantities
\begin{equation}
E \equiv f(r)\,\dot{t}, 
\qquad 
L \equiv r^2 \dot{\phi},
\label{eq:conserved}
\end{equation}
where an overdot denotes differentiation with respect to the proper time $\tau$ for timelike geodesics or with respect to the affine parameter $\lambda$ for null geodesics.  

The normalization condition for the four-velocity reads
\begin{equation}
g_{\mu\nu}\dot{x}^\mu \dot{x}^\nu = -\epsilon, 
\qquad
\epsilon = 
\begin{cases}
1, & \text{timelike},\\
0, & \text{null}.
\end{cases}
\label{eq:normalization}
\end{equation}
From Eqs.~\eqref{eq:line_element}–\eqref{eq:normalization} one obtains the radial equation of motion.

\subsection{Timelike geodesics}
For massive particles ($\epsilon=1$), the radial equation becomes
\begin{eqnarray}
\dot{r}^2 &=& E^2 - V_{\rm m}(r),\\
V_{\rm m}(r) &=& f(r)\left(1 + \frac{L^2}{r^2}\right),
\label{eq:timelike}
\end{eqnarray}
Circular orbits at $r=r_0$ are determined by the conditions $\dot{r}=0$ and $V_{\rm m}'(r_0)=0$, which yield
\begin{eqnarray}
L^2(r_0) &=& \frac{r_0^3 f'(r_0)}{2f(r_0) - r_0 f'(r_0)},\\
E^2(r_0) &=& \frac{2 f^2(r_0)}{2f(r_0) - r_0 f'(r_0)}.
\label{eq:timelike_EL}
\end{eqnarray}
The orbital frequency measured at infinity is given by
\begin{equation}
\Omega \equiv \frac{d\phi}{dt} = \frac{\dot{\phi}}{\dot{t}} 
= \sqrt{\frac{f'(r_0)}{2 r_0}}.
\label{eq:Omega}
\end{equation}
Stability of the orbit requires $V_{\rm m}''(r_0)>0$. The marginally stable circular orbit radius $r_{\rm ms}$ can be determined from
\begin{equation}
\frac{d}{dr}\!\left(\frac{r^3 f'(r)}{2f(r)-r f'(r)}\right)\bigg|_{r=r_{\rm ms}}=0.
\label{eq:isco}
\end{equation}

\subsection{Null geodesics}
\label{subsec:null_geodesics}
For photons ($\epsilon=0$) the radial equation takes the form
\begin{eqnarray}
\dot{r}^2 &=& E^2 - V(r
),\\
V(r)&=&\frac{f(r) L^2}{r^2}
\label{eq:null}
\end{eqnarray}
Introducing the impact parameter $l \equiv L/E$, one obtains
\begin{equation}
\left(\frac{dr}{d\lambda}\right)^2 = E^2\left(1 - \frac{f(r)l^2}{r^2}\right).
\label{eq:null_b}
\end{equation}
Circular photon orbits (the photon sphere) at $r=r_{\rm ph}$ satisfy
\begin{equation}
r_{\rm ph}\, f'(r_{\rm ph}) - 2 f(r_{\rm ph}) = 0,
\label{eq:photon_sphere}
\end{equation}
and the corresponding critical impact parameter is
\begin{equation}
l_{\rm ph}^2 = \frac{r_{\rm ph}^2}{f(r_{\rm ph})}.
\label{eq:bph}
\end{equation}

\subsection{Photon travel time and swept angle}
For null geodesics in the metric~\eqref{eq:line_element}, the conserved quantities~\eqref{eq:conserved} together with the radial equation of motion for photons~\eqref{eq:null_b} imply the first--order relations
\begin{eqnarray}
\frac{dt}{dr} &=& \pm \frac{1}{\,f(r)\,\sqrt{1 - \dfrac{f(r)l^2}{r^2}}\,},\\
\frac{d\phi}{dr} &=& \pm \frac{l}{\,r^2\,\sqrt{1 - \dfrac{f(r)l^2}{r^2}}\,}.
\label{eq:dt_dr_dphi_dr}
\end{eqnarray}
The upper (lower) sign corresponds to increasing (decreasing) $r$ along the ray.
The radial turning point $r_{\rm p}$, when present, is defined by
$1 - f(r_{\rm p})\,l^2/r_{\rm p}^2 = 0$.

\section{Time-like and null geodesics of double-peak potentials}\label{sec:calculations}

The metric exhibits non-monotonic behavior in certain regions of the parameter space, leading to the appearance of a secondary peak in the effective potential governing particle motion. To illustrate this property, we consider the parameter set \(M = 1\), \(Q = 0.01\), \(\alpha = -0.01\), and \(\beta = 1.02\). These values are not unique but represent one example of such qualitatively distinctive behavior.

\subsection{Horizons structure}

The structure of horizons in this spacetime depends sensitively on the choice of the parameters 
\(\alpha\) and \(\beta\), while keeping \(M\) and \(Q\) fixed. The analysis of the  metric function \(f(r)\) reveals a variety of  qualitatively distinct regimes.

\begin{figure}[!htb]
  \centering
  \includegraphics[width=0.48\textwidth]{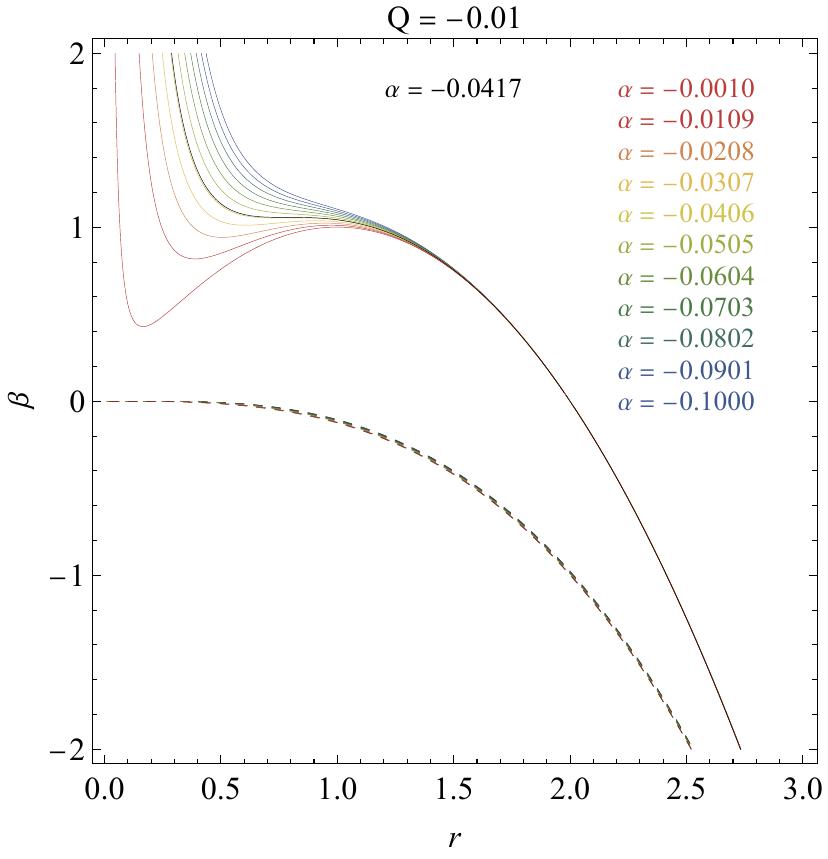}
  \caption{Contour plot illustrating the positions of horizons, i.e., values of the radial coordinate $r$ satisfying $f(r)=0$, in the parameter space $(r,\beta)$ for fixed $Q=-0.01$ and various values of the parameter $\alpha$. The black line corresponds to the critical value $\alpha \approx -0.0417$, separating the regime with multiple horizons ($\alpha<-0.0417$) from the regime with a single horizon ($\alpha>-0.0417$). Dashed lines indicate the loci where the square root in the metric function vanishes; below these curves the metric becomes imaginary and the spacetime is not defined.}
  \label{fig:horizon_a}
\end{figure}

In Fig.~\ref{fig:horizon_a} we present a contour plot showing the horizon positions in the parameter space $(r,\beta)$ for fixed $Q=-0.01$ and different values of the parameter $\alpha$. The diagram clearly demonstrates the existence of a critical value $\alpha \approx -0.0417$, marked in black. For $\alpha<-0.0417$ the spacetime admits multiple horizons, while for $\alpha>-0.0417$ only a single horizon exists. The dashed curves correspond to the loci where the square root in the definition of the metric function vanishes. For parameter values and radial coordinates below these dashed curves the metric function takes imaginary values, and the spacetime becomes ill-defined. 

\begin{figure}[!htb]
  \centering
  \includegraphics[width=0.48\textwidth]{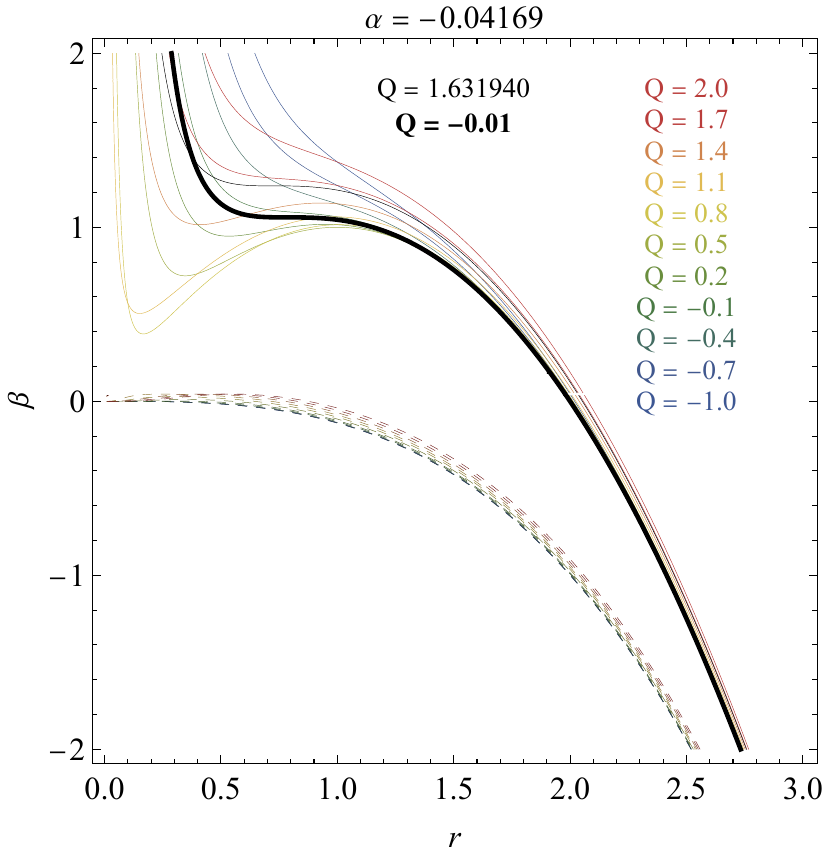}
  \caption{Contour plot of horizon positions, i.e., values of the radial coordinate $r$ satisfying $f(r)=0$, in the parameter space $(r,\beta)$ for fixed $\alpha=-0.04169$ and various values of the additional integration constant $Q$ tied to the Proca field. Two critical values of $Q$ can be identified from the diagram: $Q \approx -0.01$ and $Q \approx 1.63194$. For $-0.01<Q<1.63194$ the spacetime admits multiple horizons, while for $Q<-0.01$ and $Q>1.63194$ only a single horizon exists. Dashed lines indicate the loci where the square root in the metric function vanishes; below these curves the metric becomes imaginary and the spacetime is not defined. The special case $Q=M=1$ is not considered here.}
  \label{fig:horizon_q}
\end{figure}

In Fig.~\ref{fig:horizon_q} we present a contour plot showing the horizon positions in the parameter space $(r,\beta)$ for fixed $\alpha=-0.04169$ and different values of the additional integration constant $Q$ related to the Proca field. The plot reveals the existence of two critical values, $Q \approx -0.01$ and $Q \approx 1.63194$. For values of $Q$ between these limits, the spacetime admits multiple horizons, while for $Q<-0.01$ and $Q>1.63194$ only a single horizon remains. The dashed curves correspond to the loci where the square root in the metric function vanishes; for parameter values and radial coordinates below these curves, the metric function becomes imaginary and the spacetime is ill-defined. The case $Q=M=1$ is excluded from our considerations.

\begin{figure}[!htb]
  \centering
  \includegraphics[width=0.95\linewidth]{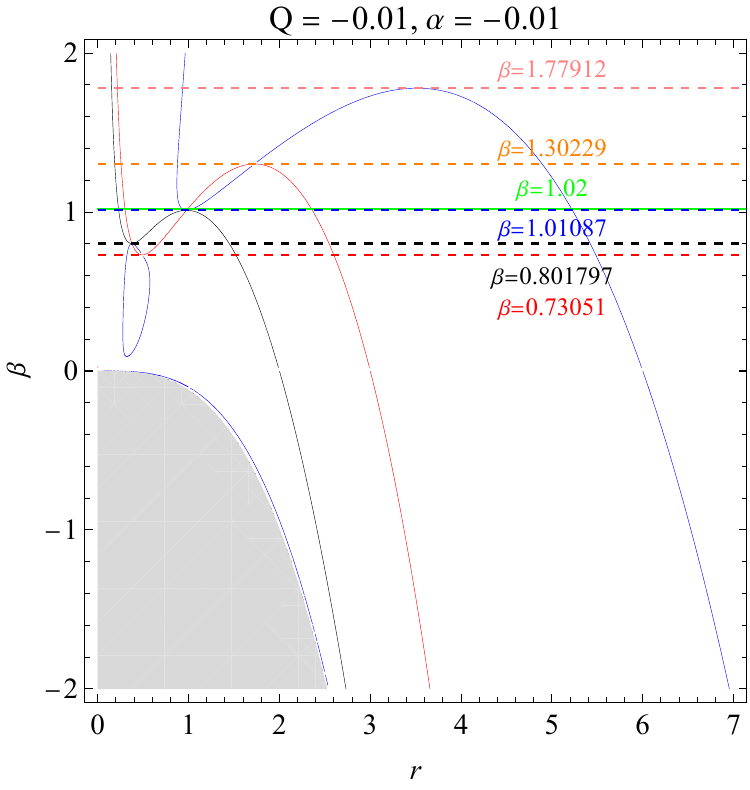}\\[1ex]
  \includegraphics[width=0.95\linewidth]{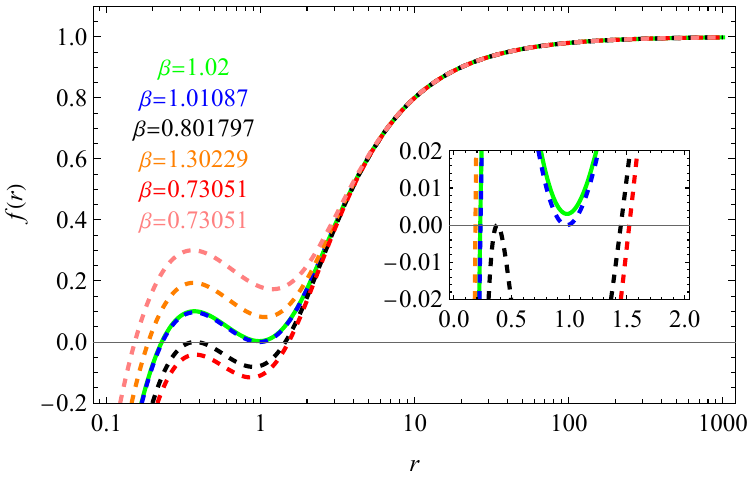}
  \caption{\textbf{Top panel:} dependence of characteristic radii on the parameter $\beta$ 
  for fixed values $M=1$, $Q=-0.01$ and $\alpha=-0.01$. 
  The black curve shows the location of the black hole horizon. 
  The red curve corresponds to the radii of photon circular orbits, while the blue curve 
  indicates the radii of marginally stable circular orbits for massive particles. 
  The grey-shaded region marks the values of $\beta$ for which the metric function $f(r)$ 
  becomes imaginary due to the square root in its definition. 
  Horizontal reference lines at $\beta=0.801797$ and $\beta=1.01087$ indicate 
  the approximate boundaries of the multi-horizon regime. 
  For $\beta$ values slightly above $\beta=1.01087$ (illustrated by the green 
  reference line at $\beta=1.02$), the spacetime again possesses a single horizon 
  of smaller radius. This transition gives rise to distinct features in particle 
  dynamics and optical phenomena. 
  \textbf{Bottom panel:} the metric function $f(r)$ as a function  of the radial coordinate $r$ 
  for selected parameter values $M, Q, \alpha, \beta$.  
  The positions of the horizons are given by the zeros of $f(r)$.}
  \label{fig:radii_b}
\end{figure}

In Fig.~\ref{fig:radii_b} (top panel) we show the dependence of the event horizon radius together with the radii of photon circular orbits and marginally stable circular orbits for massive particles on the parameter $\beta$, for fixed $M=1$, $Q=-0.01$, and $\alpha=-0.01$. The multi-horizon regime is bounded approximately by $\beta=0.801797$ and $\beta=1.01087$. For $\beta$ values slightly above $\beta=1.01087$ (for example $\beta=1.02$), the spacetime again admits a single horizon of smaller radius. This transition gives rise to distinct features in particle dynamics and optical phenomena. 

The diagram further demonstrates that a double-peak structure of the effective potential appears both for photons and massive particles. For photons, additional circular orbits exist in the interval $0.73051 \lesssim \beta \lesssim 1.30229$. However, only the range $\beta>1.01087$ is physically relevant, since for smaller values the additional photon orbits are hidden either inside the non-static region beneath the outer horizon or inside the inner static region, and thus are inaccessible to an external observer. For massive particles, the effective potential develops more than one marginally stable circular orbit in the interval $1.01087 \lesssim \beta \lesssim 1.77912$, which significantly affects the orbital dynamics in this parameter range. The bottom panel of the figure ~\ref{fig:radii_b} illustrates the corresponding behavior of the metric function $f(r)$ for representative values of \(\beta\), with the positions of the horizons given by its zeros.

\begin{figure}[!htb]
  \centering
  \includegraphics[width=0.90\linewidth]{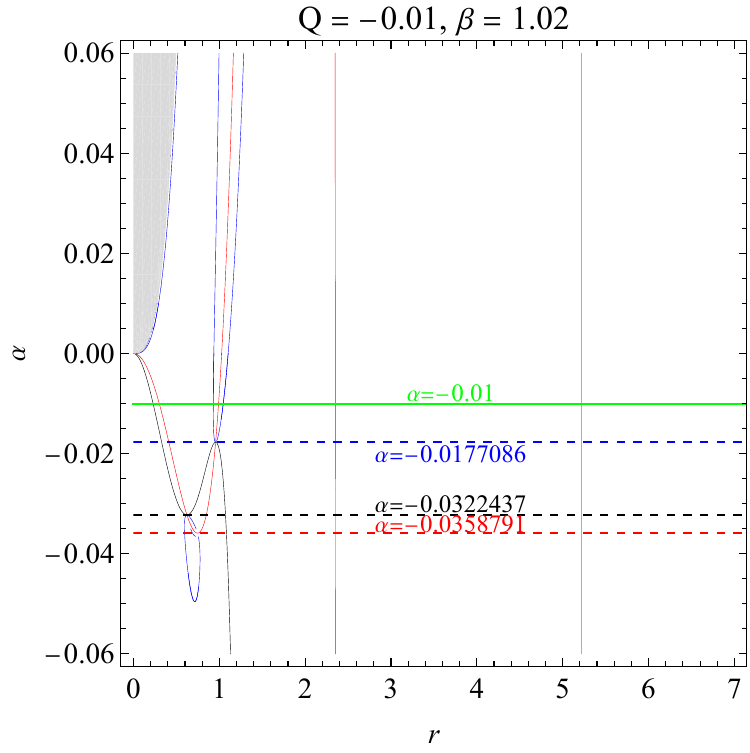}\\[1ex]
  \includegraphics[width=0.90\linewidth]{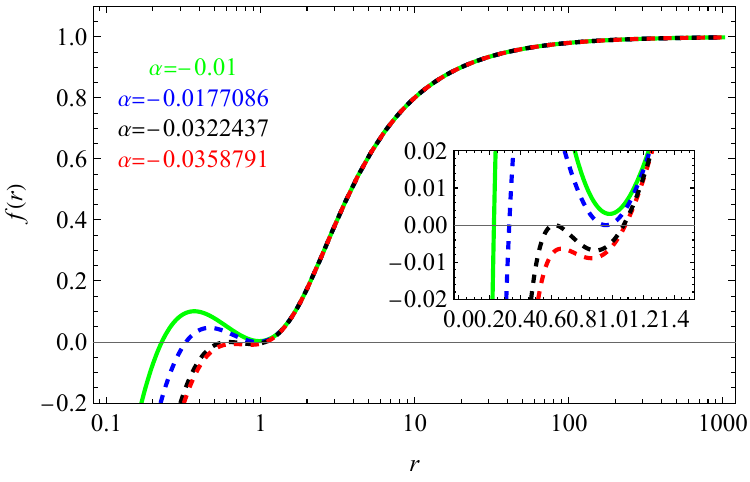}
  \caption{\textbf{Top panel:} dependence of characteristic radii on the parameter $\alpha$
  for fixed $M=1$, $Q=-0.01$ and $\beta=1.02$.
  The black curve shows the location of the black hole horizon.
  The red curve corresponds to the radii of photon circular orbits, while the blue curve 
  indicates the radii of marginally stable circular orbits for massive particles. 
  For $\alpha>0$ the spacetime does not admit any horizons; however, it cannot
  be described as a naked singularity either, since in the grey-shaded region
  the metric function $f(r)$ becomes imaginary and therefore undefined.
  Horizontal reference lines at $\alpha=-0.0322437$ and $\alpha=-0.0177086$
  indicate the parameter range where the spacetime possesses multiple horizons,
  while the line at $\alpha=-0.01$ marks a representative value chosen as an
  example for further calculations. An additional horizontal reference line at 
  $\alpha=-0.0358791$ corresponds to the minimal value of $\alpha$ at which 
  more than one photon orbit appears; nevertheless, the physically interesting case is 
  only $\alpha>-0.0177086$, since in this regime the additional photon circular 
  orbits and marginally stable orbits of massive particles lie outside the outer horizon 
  and are thus accessible to an external observer. 
  \textbf{Bottom panel:} the metric function $f(r)$ as a function of the radial coordinate $r$ 
  for selected parameter values $M, Q, \alpha, \beta$.  The behavior of the function at
  large radii shows that $f(r)\to 1$, which corresponds to an asymptotically flat spacetime.}
  \label{fig:radii_a}
\end{figure}

In Fig.~\ref{fig:radii_a} (top panel) we show the dependence of the event horizon radius, together with the radii of photon circular orbits and marginally stable circular orbits for massive particles, on the parameter $\alpha$, for fixed $M=1$, $Q=-0.01$, and $\beta=1.02$. The multi-horizon regime occurs for $-0.0322437 \lesssim \alpha \lesssim -0.0177086$. A representative value $\alpha=-0.01$ lies outside this interval and corresponds to a single-horizon configuration. For $\alpha>0$, no horizons are present, although the geometry cannot be classified as a naked singularity since the metric function becomes imaginary in the grey-shaded region due to the square root in its definition. An additional reference value, $\alpha=-0.0358791$, corresponds to the minimal $\alpha$ at which more than one photon circular orbit appears. However, the physically most relevant case is $\alpha>-0.0177086$, since in this regime the additional photon circular orbits and marginally stable orbits of massive particles lie outside the outer horizon and are thus observable from infinity. The bottom panel shows the metric function $f(r)$ for representative and boundary values of $\alpha$, with horizons given by the zeros of $f(r)$.

\subsection{Timelike circular geodesics} 

\begin{figure}[h]
  \centering
  \includegraphics[width=0.45\textwidth]{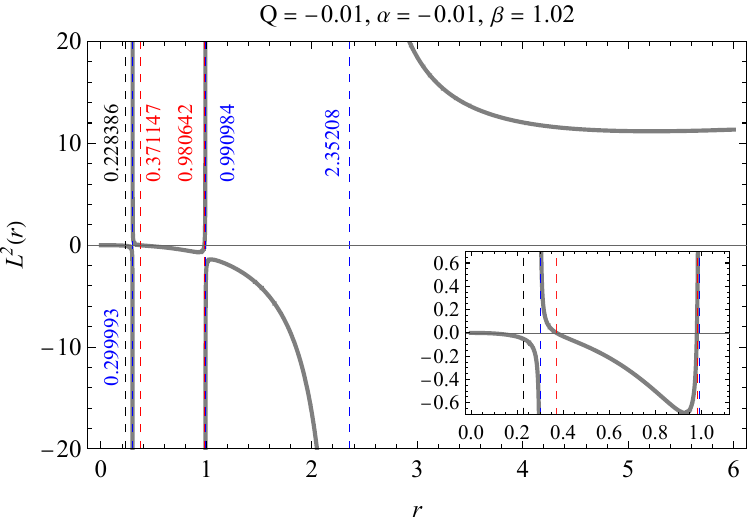}
  \includegraphics[width=0.45\textwidth]{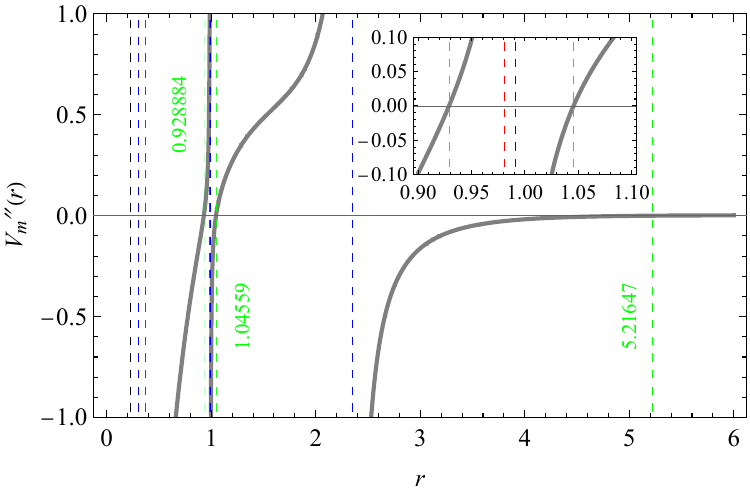}
  \caption{Radial profiles of $L^2(r)$ and $V_{\rm m}''(r)$ with 
  fixed parameters $M=1$, $Q=-0.01$, $\alpha=-0.01$, $\beta=1.02$. 
  \textbf{Top panel:} squared specific angular momentum $L^2$. 
  Vertical blue and red dashed lines indicate characteristic radii where $L^2$ either vanishes or diverges, 
  delimiting the intervals in which circular orbits with $L^2>0$ can exist. 
  The black dashed line marks the black hole horizon. 
  \textbf{Bottom panel:} second derivative of the effective potential, 
  $V_{\rm m}''(r)$. The sign of this function determines orbital stability: 
  $V_{\rm m}''(r)>0$ for stable orbits and $V_{\rm m}''(r)<0$ for unstable ones. 
  Green dashed vertical lines show the radii where $V_{\rm m}''(r)=0$. 
  The horizon position (black dashed) and the characteristic radii where $L^2$ vanishes 
  or diverges (red and blue dashed lines) coincide with those displayed in the top panel.}
  \label{fig:l2_circular}
\end{figure}

In order to characterize the orbital structure of massive particles, we focus on the conditions for the existence and stability of circular timelike geodesics. For the representative parameter set ($M=1$, $Q=-0.01$, $\alpha=-0.01$, $\beta=1.02$), the relevant quantities are displayed in Fig.~\ref{fig:l2_circular}.

The upper panel in Fig.~\ref{fig:l2_circular} shows the radial dependence of the squared specific angular momentum $L^2(r)$. Vertical red dashed lines indicate radii where $L^2$ vanishes, while blue dashed lines mark its divergences. These characteristic points delimit the allowed regions in which circular orbits with $L^2>0$ can exist. The black dashed line corresponds to the event horizon, setting the innermost boundary for circular motion.

The lower panel in Fig.~\ref{fig:l2_circular} presents the second derivative of the effective potential, $V_{\rm m}''(r)$. The sign of this quantity governs orbital stability: $V_{\rm m}''(r)>0$ signals stable orbits, whereas $V_{\rm m}''(r)<0$ corresponds to instability. The transition radii where $V_{\rm m}''(r)=0$ (green dashed lines) coincide with changes of stability. Comparison with the $L^2(r)$ profile shows that stable circular motion is possible only within intervals where both $L^2>0$ and $V_{\rm m}''(r)>0$ hold simultaneously.

Overall, the combined analysis of $L^2(r)$ and $V_{\rm m}''(r)$ reveals a nontrivial structure of the circular orbit spectrum, with alternating stable and unstable regions outside the horizon. The presence of multiple intervals for the existence of circular timelike orbits suggests a rich orbital dynamics specific to this spherically symmetric black-hole solution. In particular, for the chosen parameters the stable circular orbits are confined to two separate radial domains:
\[
r \in (0.980642,\,0.990984) 
\quad \text{and} \quad 
r \in (5.21647,\,\infty).
\]

\subsection{Effective potentials for timelike geodesics}

\begin{figure}[h]
  \centering
  \includegraphics[width=0.45\textwidth]{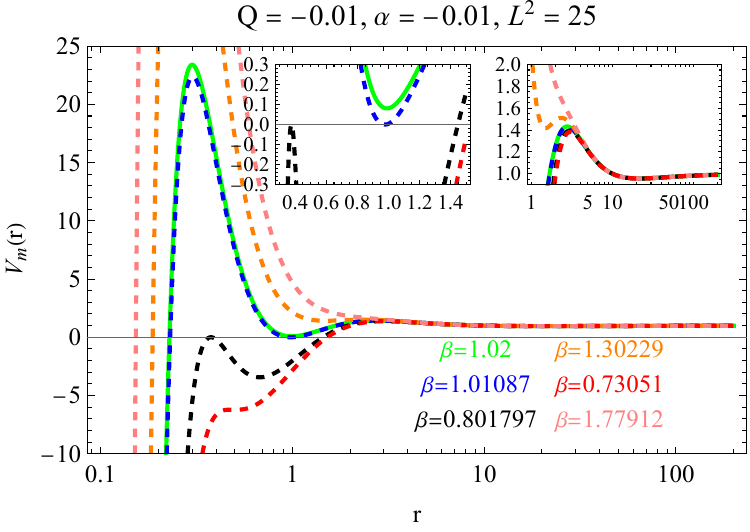}
  \includegraphics[width=0.45\textwidth]{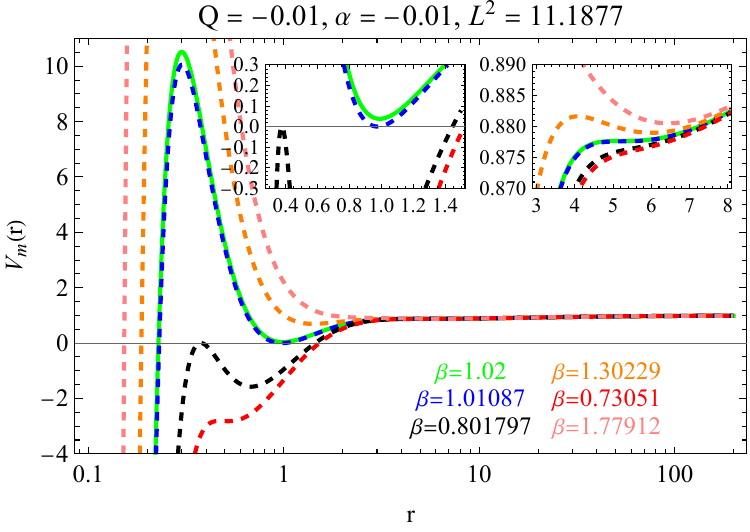}
  \caption{Effective potential for timelike geodesics, $V_m(r)$, for fixed
  $M=1$, $Q=-0.01$, $\alpha=-0.01$, and two values of $L^2$. 
  The curves correspond to 
  $\beta=0.73051,\;0.801797,\;1.01087,\;1.02,\;1.30229,\;1.77912$, chosen in
  accordance with the reasons outlined earlier (multi–horizon bounds and the
  intervals where additional photon or massive–particle orbits occur). 
  Circular orbits occur at extrema of $V_m$ (minima: stable; maxima: unstable). 
  \textbf{Top panel:} $L^2=25$. For $\beta\gtrsim1.01087$ a double–peak structure
  appears outside the horizon (two minima and two maxima), allowing two families
  of bound orbits; within/below the window only a single maximum–minimum pair (or
  a monotonic profile) remains, permitting at most one stable orbit. 
  \textbf{Bottom panel:} $L^2=11.1877$. For $\beta=1.02$ this value of $L^2$ is
  critical, marking the marginal stability of the outer orbit. For larger
  $\beta$ up to $\beta=1.77912$ the double–peak structure persists, while at
  $\beta=1.77912$ the second peak disappears completely. For $\beta\le1.01087$
  the potential is monotonic above the horizon and no stable circular orbits
  exist at this $L^2$.}
  \label{fig:Vm}
\end{figure}

To explore the orbital dynamics of massive particles, we analyze the effective
potential $V_m(r)$ for a set of $\beta$ values selected in accordance with the
reasons discussed in the previous sections, namely their role in delimiting the
multi–horizon regime and in determining the existence of additional photon and
massive–particle orbits. Figure~\ref{fig:Vm} shows $V_m(r)$ for fixed
$M=1$, $Q=-0.01$, $\alpha=-0.01$, and two representative values of $L^2$.
Circular orbits occur at the extrema of $V_m(r)$ ($dV_m/dr=0$), with minima
corresponding to stable and maxima to unstable orbits.

As illustrated in the top panel of Fig.~\ref{fig:Vm} for $L^2=25$, 
when $\beta$ is slightly above the upper boundary of the multi–horizon regime
($\beta \gtrsim 1.01087$, e.g., $\beta=1.02,\,1.30229,\,1.77912$), 
the potential develops a pronounced double–peak structure outside the horizon 
(two minima separated by two maxima), enabling two distinct families of bound 
orbits. In contrast, for $\beta$ within or below the multi–horizon window
($\beta=1.01087,\,0.801797,\,0.73051$) the profile exhibits at most a single
maximum–minimum pair above the horizon—and for the smallest $\beta$ it becomes
monotonic—so that at most one stable circular orbit is available at this $L^2$.

The bottom panel of Fig.~\ref{fig:Vm} corresponds to $L^2=11.1877$. 
This value of $L^2$ is critical for $\beta=1.02$, where the outer circular orbit 
becomes marginally stable. For larger values of $\beta$ (e.g.\ $\beta=1.30229$), 
the potential still exhibits a double–peak structure outside the horizon, so that 
two families of circular orbits persist. This continues up to the limiting value 
$\beta=1.77912$, at which the second peak disappears completely and the potential 
reduces to a single–well form. For $\beta\leq 1.01087$ the potential is monotonic 
above the horizon, and no stable circular orbits exist for this $L^2$.

\subsection{Effective potentials for null geodesics}

\begin{figure}[h]
  \centering
  \includegraphics[width=0.45\textwidth]{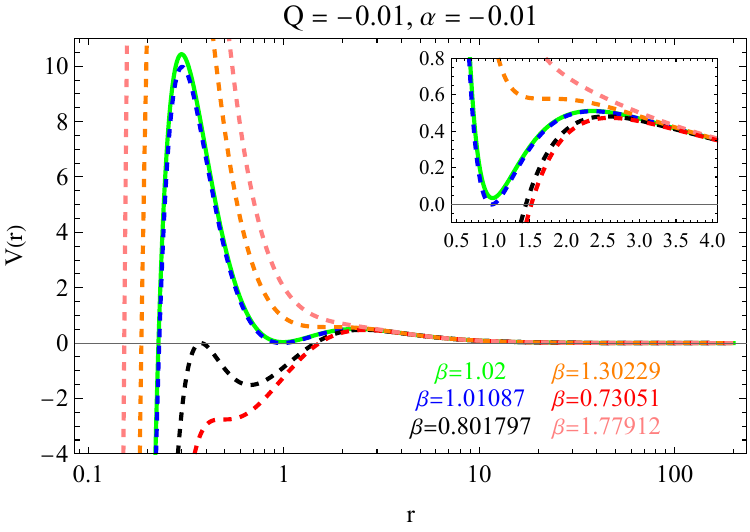}
  \caption{Effective potential for photon (null) geodesics, $V(r)$, as a function 
  of radial coordinate $r$ for fixed $Q=-0.01$ and $\alpha=-0.01$. 
  The curves correspond to 
  $\beta=0.73051,\;0.801797,\;1.01087,\;1.02,\;1.30229,\;1.77912$, 
  chosen in accordance with the reasons outlined earlier. 
  For $\beta=0.73051,\;0.801797$ and $1.01087$, the potential exhibits a single 
  maximum outside the horizon, while the inner peak is hidden beneath the outer 
  horizon or inside the inner static region, and is therefore inaccessible to an 
  external observer. For $\beta=1.02$, a double–peak structure arises, implying 
  the existence of two unstable and one stable circular photon orbit; the latter 
  is associated with a potential well supporting bound ``rosette'' photon 
  trajectories. For $\beta=1.30229$, the second outer peak disappears, marking a 
  critical case where a marginally stable photon orbit occurs. For $\beta=1.77912$, 
  above the critical value, the double–peak structure does not form at all, and only 
  a single unstable photon orbit remains.}
  \label{fig:V_photon}
\end{figure}

To investigate the properties of null geodesics, we analyze the effective potential 
$V(r)$ for photons. Figure~\ref{fig:V_photon} displays representative profiles of 
$V(r)$ for fixed $M=1$, $Q=-0.01$, $\alpha=-0.01$, and a set of $\beta$ values 
selected in accordance with the reasons discussed earlier, i.e.\ the boundaries of 
the multi–horizon regime and the interval where additional photon orbits may occur. 
The structure of the potential determines the existence and stability of 
circular photon orbits: maxima correspond to unstable photon spheres, while minima 
indicate possible stable photon orbits.

For $\beta=0.73051,\;0.801797$ and $\beta=1.01087$, the potential exhibits a single 
maximum outside the horizon, while the inner peak is hidden either inside the non–static 
region beneath the outer horizon or inside the inner static region, and is therefore 
inaccessible to an external observer. In these cases, only one unstable photon orbit 
is present, corresponding to the usual photon sphere in spherically symmetric 
black–hole spacetimes.  

For $\beta=1.02$, the potential develops a richer structure: two maxima appear 
above the horizon, separated by a minimum. This configuration implies the existence 
of two distinct unstable photon spheres and, in addition, one stable circular photon 
orbit at the position of the minimum. In the vicinity of this stable orbit, an isolated 
potential well forms, giving rise to bound photon trajectories of a ``rosette'' type, 
which remain confined and disconnected from observers at infinity.  

For $\beta=1.30229$, the second (outer) peak disappears, so that this value represents 
a critical case where a marginally stable photon orbit occurs. For $\beta=1.77912$, 
which lies above this critical value, the double–peak structure does not form at all, 
and remains only a single unstable photon orbit.

\section{Effects in the presence of Proca hair}

Having established the general equations of motion in Sec.~\ref{sec:eq_motion},
we now turn to their application in the spacetime of a black hole with
Primary Proca hair. Our goal here is to analyze the purely geometrical
properties of photon trajectories that determine the observable image
features (we use standard techniques here, see e.g. \cite{Stuchlik:2019apj}). 

For concreteness, we adopt a configuration in which both the source
and the observer are located on a sphere of radius $r=200M$.
The source is assumed to have a finite angular extent of 
$\Delta\psi = 0.2$ .
The observer, the black hole center, and the source lie in the same plane,
but not along the same straight line. The observer is placed at
\[
(r_o,\theta_o,\phi_o) = (200M,\,\pi/2,\,0),
\]
while the source is located at
\[
(r_s,\theta_s,\phi_s) = (200M,\,\pi/2 -0.25,\, \pi-0.02).
\]
This setup allows us to study how the finite source size is mapped into
the observer’s sky by null geodesics. Within this framework we compute
the photon travel time, the swept angle, and the corresponding image
characteristics for different orders of trajectories circling the black hole.

\subsection{Time delay}

We evaluate the photon travel time as a function of the impact parameter~$l$. 
Both the source and the observer are located on the sphere of radius $r=200M$, 
with the observer at $(r_o,\theta_o,\phi_o)=(200M,\,\pi/2,\,0)$ 
and the source corresponding to a finite patch of the same spherical surface. 
In this configuration only photon trajectories with a turning point connect the source and the observer.  

The turning point $r_{\rm p}$ is defined by the condition
\begin{equation}
1 - \frac{f(r_{\rm p})\,l^2}{r_{\rm p}^2} = 0,
\label{eq:turning_point}
\end{equation}
which follows from the radial equation of motion $\dot r^2=0$. 
The coordinate travel time is thus determined by summing the ingoing and outgoing contributions, through the relation \(dt/dr\) given in Eq.~\eqref{eq:dt_dr_dphi_dr}, leading to
\begin{equation}
\Delta t(l) = \int_{r_s}^{r_{\rm p}} 
\frac{dr}{\,f(r)\,\sqrt{1 - \dfrac{f(r)l^2}{r^2}}\,} 
+ \int_{r_o}^{r_{\rm p}} 
\frac{dr}{\,f(r)\,\sqrt{1 - \dfrac{f(r)l^2}{r^2}}\,}.
\label{eq:ToF_turning}
\end{equation}

\begin{figure}[h]
  \centering
  \includegraphics[width=0.45\textwidth]{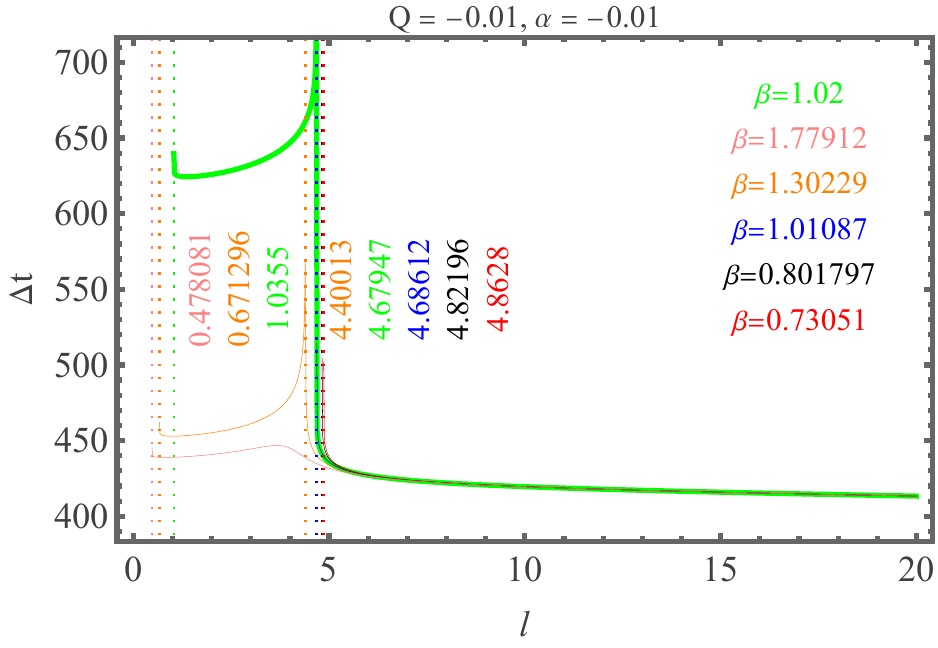}
  \caption{Photon travel time $\Delta t$ as a function of the impact parameter $l$ 
  for $Q=-0.01$ and $\alpha=-0.01$, shown for several values of the parameter 
  $\beta$. Both source and observer are located on a sphere of radius $r=200M$. 
  For large $l$ the travel time corresponds to the weak--lensing regime. 
  As $l$ approaches the critical values associated with photon spheres, $\Delta t$ diverges, 
  reflecting the fact that photons spend an arbitrarily long coordinate time orbiting near the unstable circular orbit before reaching the observer. 
  The vertical lines mark the critical impact parameters $l_{\rm ph}$ of the photon orbits, 
  whose numerical values are indicated on the plot for each curve.}
  \label{fig:TimeDel}
\end{figure}

Figure~\ref{fig:TimeDel} shows the dependence of $\Delta t$ on the impact parameter $l$. 
For large $l$ the delay is small, corresponding to the weak--lensing regime. 
Near the two critical values of $l$, the photon spends an increasingly long time 
orbiting the black hole before reaching the observer, leading to a logarithmic divergence 
of $\Delta t(l)$.

\subsection{Image angular size}

Next we analyze the swept angle as a function of the impact parameter~$l$. 
For a finite--distance source and observer, both placed on the sphere of radius $r=200M$, 
the total azimuthal angle swept by the photon trajectory is obtained from the relation 
$d\phi/dr$ in Eq.~\eqref{eq:dt_dr_dphi_dr}, leading to
\begin{equation}
\Delta \phi(l) = \int_{r_s}^{r_{\rm p}} 
\frac{l\,dr}{\,r^2\,\sqrt{1 - \dfrac{f(r)l^2}{r^2}}\,} 
+ \int_{r_o}^{r_{\rm p}} 
\frac{l\,dr}{\,r^2\,\sqrt{1 - \dfrac{f(r)l^2}{r^2}}\,},
\label{eq:phi_turning}
\end{equation}
where $r_s=r_o=200M$ and $r_{\rm p}$ denotes the radial turning point, 
defined in analogy with Eq.~\eqref{eq:turning_point} considered earlier.  

\begin{figure}[h]
  \centering
  \includegraphics[width=0.45\textwidth]{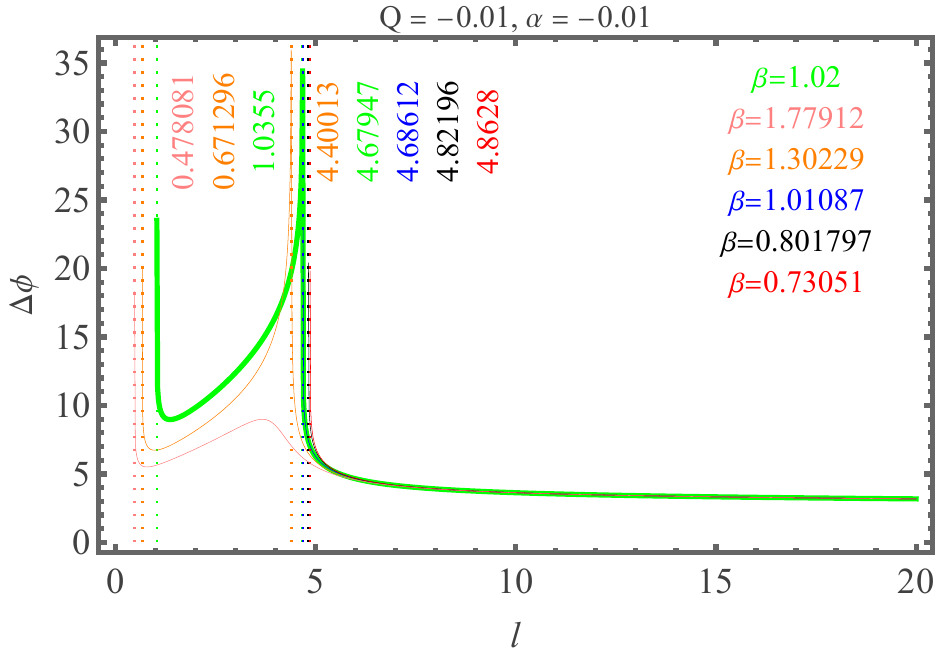}
  \caption{Swept angle $\Delta\phi$ as a function of the impact parameter $l$ 
  for $Q=-0.01$ and $\alpha=-0.01$, shown for several values of the parameter 
  $\beta$. For large $l$ the angle decreases, corresponding to the weak--lensing 
  regime. As $l$ approaches the critical values associated with photon spheres, 
  $\Delta\phi$ diverges, reflecting the fact that photons can orbit an arbitrary 
  number of times around the black hole before escaping. The vertical lines mark 
  the critical impact parameters $l_{\rm ph}$ of the photon orbits, with their 
  numerical values indicated on the plot for each curve.}
  \label{fig:DefAng}
\end{figure}

The resulting dependence $\Delta \phi(l)$ is shown in Fig.~\ref{fig:DefAng} 
for several values of the parameter $\beta$, with $Q=-0.01$ and $\alpha=-0.01$.  
For large $l$ the swept angle decreases, corresponding to the weak--lensing regime. 
As $l$ approaches the critical values associated with photon spheres, $\Delta\phi(l)$ 
diverges, since photons may orbit the black hole an arbitrary number of times 
before escaping to the observer. The vertical lines in Fig.~\ref{fig:DefAng} 
indicate the critical impact parameters $l_{\rm ph}$ of the photon orbits, 
with their numerical values displayed on the plot for each curve. 

\subsection{Critical impact parameters}\label{subsec:critical_impact}
The conditions defining the photon sphere and its critical impact parameter
have been derived in Sec.~\ref{subsec:null_geodesics}, see Eqs.~(\ref{eq:photon_sphere}) 
and~(\ref{eq:bph}). These expressions follow from the radial equation of null 
geodesics [Eq.~(\ref{eq:null_b})] under the requirements $\dot r=0$ and
$dV/dr=0$. Here we apply these general formulas to determine explicitly the
photon circular orbits in our spacetime and classify their observational relevance. 

\begin{table*}[t]
\centering
\caption{Photon-orbit color key and relevance for the shadow 
(see Fig.~\ref{fig:shadow_lb}).}
\begin{tabular}{llll}
\hline\hline
Color / Style & Orbit (regime) & Accessibility & Shadow relevance \\
\hline
Orange, solid / solid & Single photon orbit (one horizon) & Outside horizons & Defines shadow radius \\
Black, solid / solid  & External unstable (multi--horizon) & Outside horizons & Defines shadow radius \\
Brown, dotted / no line & --- (multi--horizon) & NS region & No \\
Green, solid / dotted & Internal unstable (multi--horizon) & Inaccessible (IS region) & No \\
Pink, solid / dashed  & External unstable (one horizon) & Accessible but not limiting & No \\
Purple, solid / dash--dotted & Internal stable (one horizon) & Inaccessible for DO & No \\
Red, solid / solid    & Internal unstable (one horizon) & Outside horizons & Defines shadow radius \\
Blue, solid / solid  & Single photon orbit (one horizon) & Outside horizons & Defines shadow radius \\
\hline\hline
\end{tabular}
\footnotetext{NS = non--static region; IS = inner--static region; DO = distant observer. “Accessible” means 
outside the outer horizon and outside the non--static region.}
\label{tab:shadow_orbits}
\end{table*}

\begin{figure}[!htb]
  \centering
  \includegraphics[width=0.45\textwidth]{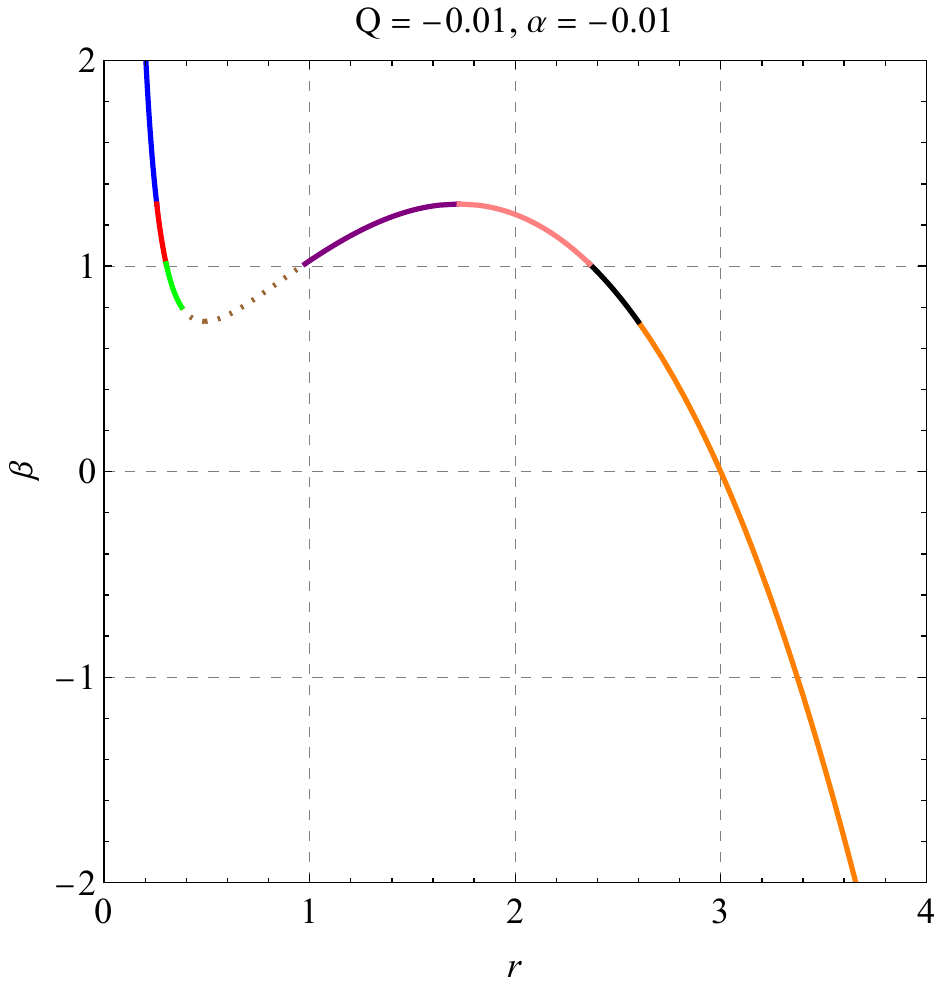}
  \includegraphics[width=0.45\textwidth]{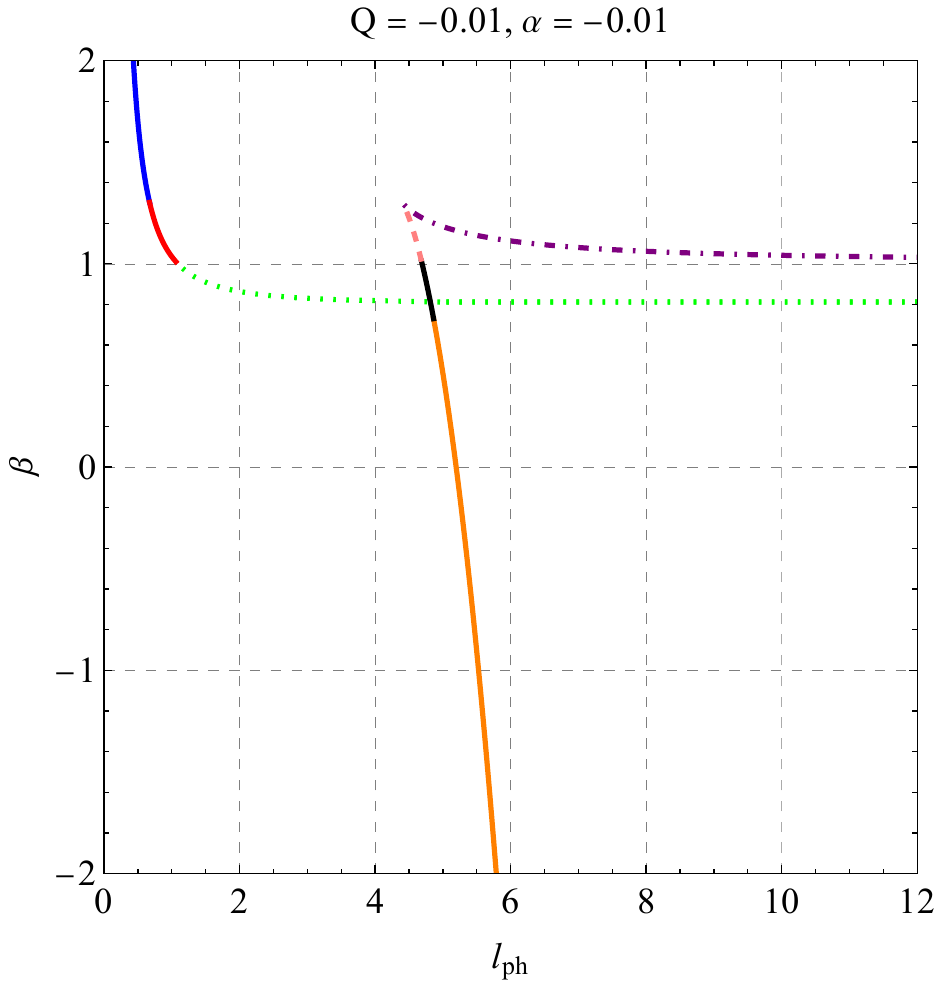}
  \caption{Photon circular orbits and corresponding critical impact parameters 
  for $M=1$, $Q=-0.01$, and $\alpha=-0.01$. 
  \textbf{Top panel:} radius of the photon orbit as a function of the parameter 
  $\beta$. Depending on $\beta$, either a single photon orbit exists or multiple 
  photon orbits appear simultaneously. Additional peaks in effective potential arise in the interval 
  $0.73051 \lesssim \beta \lesssim 1.30229$, but only for $\beta>1.01087$ 
  they are located outside the outer horizon and thus accessible to an 
  external observer.
  \textbf{Bottom panel:} the corresponding critical impact parameters $l_{\rm ph}$, 
  which determine the apparent shadow of the black hole. 
  The color coding of the curves corresponds to the classification 
  summarized in Table~\ref{tab:shadow_orbits}.}
  \label{fig:shadow_lb}
\end{figure}

We determine the critical impact parameters associated with photon circular orbits.
The two panels of Fig.~\ref{fig:shadow_lb}, together with the summary in 
Table~\ref{tab:shadow_orbits}, provide a complete classification of photon 
circular orbits in the considered spacetime. The upper panel shows how the 
radii of photon orbits depend on the parameter $\beta$, while the lower panel 
illustrates the corresponding critical impact parameters $l_{\rm ph}$ that 
determine the shadow boundary. The color coding consistently distinguishes 
between the different regimes, indicating whether only one or multiple photon 
orbits are present. 

As summarized in Table~\ref{tab:shadow_orbits}, not all photon orbits are 
physically relevant for an external observer. Only those located outside the 
outer horizon and outside the non--static region contribute to the observable 
shadow. In the multi--horizon regime ($0.801797 \leq \beta \leq 1.01087$), the 
black line marks the external unstable photon orbit, which defines the shadow 
radius, while the additional inner orbits remain hidden. In the interval 
$1.01087 \leq \beta \leq 1.30229$, several photon orbits coexist, but the red 
branch sets the shadow radius. For $\beta$ outside this interval, only a single 
orbit exists (orange or blue curves), and its critical impact parameter uniquely 
determines the shadow size.

\subsection{Shadows}

In this subsection, we compute the angular size of the shadow produced by a Proca black hole and compare it with the Schwarzschild case. As a reference, we adopt the black hole in the M87 galaxy.

\begin{figure}[!htb]
  \centering
  \includegraphics[width=0.45\textwidth]{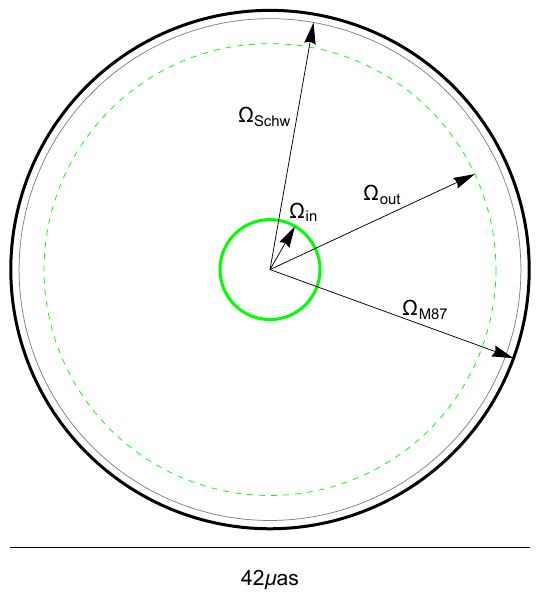}
  \caption{Shadow radii of a black hole with Primary Proca Hair, computed for 
  $Q=-0.01$, $\alpha=-0.01$, and $\beta=1.02$. 
  The curves are shown as circles in the observer’s sky, with the radius expressed 
  in microarcseconds for a black hole with the mass and distance of M87*. 
  The green circles denote the shadows corresponding to the inner and outer circular 
  photon orbits of the Primary Proca Hair black hole 
  ($\Omega_{\rm in}=3.9349\,\mu$as, $\Omega_{\rm out}=17.782\,\mu$as). 
  For comparison, the grey circle shows the shadow radius of a Schwarzschild 
  black hole with the same mass and distance 
  ($\Omega_{\rm Schw}=19.7453\,\mu$as), while the black circle corresponds to the average
  measured shadow size of M87* ($\Omega_{87}=21\,\mu$as) as reported by the 
  EHT.}
  \label{fig:Shadow}
\end{figure}

The apparent shadow boundary is set by null rays with the critical impact parameter $l_{\rm ph}$ of the relevant photon circular orbit. 
For a distant observer at distance $D$, the angular radius of the shadow is
\begin{equation}
\Omega \;=\; \frac{l_{\rm ph}}{D} \;=\; \Big(\frac{l_{\rm ph}}{M}\Big)\,\theta_M,
\qquad 
\theta_M \equiv \frac{GM}{Dc^2},
\label{eq:Omega_general}
\end{equation}
where $\theta_M$ is the angular size of one gravitational radius.

For M87*, EHT (Event Horizon Telescope) analyzes give $\theta_M = GM/(Dc^2) = 3.8 \pm 0.4~\mu{\rm as}$ when adopting $D=16.8\pm0.8$~Mpc and $M \simeq 6.5\times10^9\,M_\odot$, a commonly used value for M87 in the Virgo cluster distance scale\citep{EHT2019VI,Akiyama2024_persistent}, one has
\begin{equation} 
\Omega_{\rm Schw} \approx \sqrt{27}\times 3.8~\mu{\rm as} \approx 19.7453~\mu{\rm as},
\label{eq:Omega_Schw_164}
\end{equation}
for a Schwarzschild black hole (since $l_{\rm ph}/M=\sqrt{27}$), which matches the value used for the grey reference circle in Fig.~\ref{fig:Shadow}.

In case of Proca black hole, the critical curve(s) $l_{\rm ph}(\beta)$ were obtained in Sec.~\ref{subsec:critical_impact}. 
Using Eq.~\eqref{eq:Omega_general} with the same M87* scaling yields the two angular radii shown by the green circles in Fig.~\ref{fig:Shadow}:
\begin{equation}
\Omega_{\rm in} \;=\; \Big(\frac{l_{\rm ph,in}}{M}\Big)\,\theta_M,
\qquad
\Omega_{\rm out} \;=\; \Big(\frac{l_{\rm ph,out}}{M}\Big)\,\theta_M,
\end{equation}
which evaluate, for $l_{\rm ph,in}=1.0355$ and $l_{\rm ph,out}= 4.67947$  ($Q=-0.01$, $\alpha=-0.01$, and $\beta=1.02$), to
\[
\Omega_{\rm in}=3.9349~\mu{\rm as}, 
\qquad 
\Omega_{\rm out}=17.782~\mu{\rm as}.
\]

\begin{figure}[!htb]
  \centering
  \includegraphics[width=0.75\columnwidth]{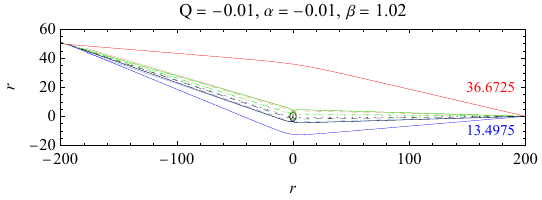}\\[1.2ex]
  \includegraphics[width=0.72\columnwidth]{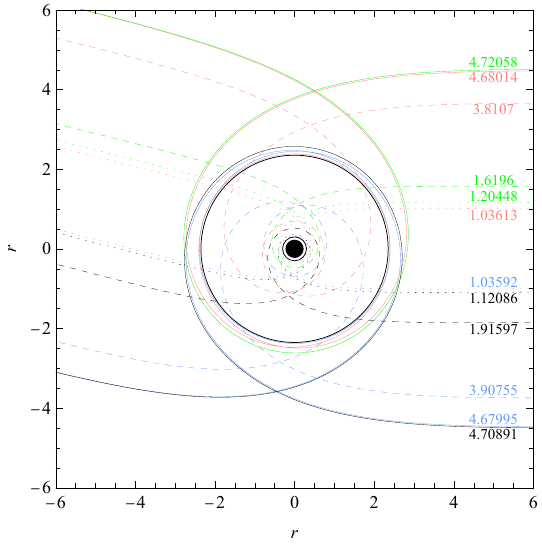}\\[1.2ex]
  \includegraphics[width=0.72\columnwidth]{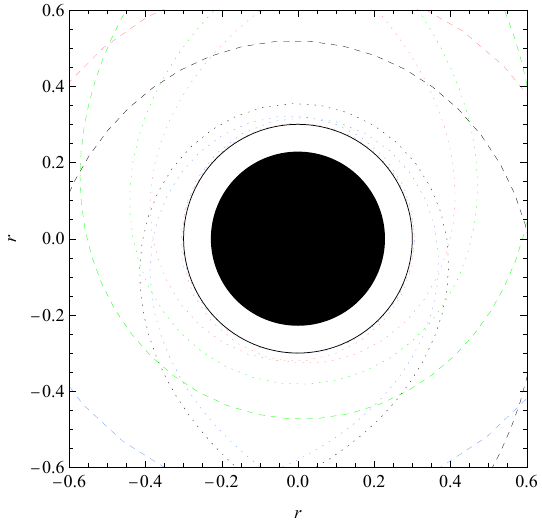}
  \caption{Photon trajectories (null geodesics) in the spacetime of a Proca black hole 
  with $Q=-0.01$, $\alpha=-0.01$, and $\beta=1.02$, for the source–observer 
  configuration. 
  \textbf{Top panel:} photon paths in the central plane connecting the source 
  $(\theta_s,\phi_s)$ and the observer $(\theta_o,\phi_o)$, both located 
  on a sphere of radius $200M$. 
  \textbf{Middle panel:} magnified view of the central region, highlighting the 
  winding trajectories near the black hole. The larger black circle at $r=2.35208$ 
  corresponds to the outer photon circular orbit; although it does not define 
  the shadow (since photon trajectories below it can still escape to infinity), 
  its presence leads to additional possible source positions. 
  \textbf{Bottom panel:} further zoom into the innermost region, resolving the 
  structure of higher–order trajectories. The black filled circle at $r=0.228386$ 
  denotes the event horizon, while the surrounding black circle at $r=0.299993$ 
  indicates the inner photon circular orbit, which determines the shadow boundary. 
  The color coding of the curves, impact parameters $l$, swept angles $\Delta\phi$, and travel times $\Delta t$ 
  corresponding to the plotted trajectories are summarized in 
  Table~\ref{tab:time_del}.}
  \label{fig:PhGeo}
\end{figure}

\subsection{Gravitational lensing}\label{subsec:grav_lens}

In this section, we focus on the illustration of gravitational lensing for the chosen 
source–observer configuration at $r_s=r_o=200M$. 

As a first step, we analyze the photon trajectories (null geodesics) that connect the source and the observer 
and are responsible for the formation of multiple images of different orders. 
These trajectories, connecting the center of the source with the observer, corresponding to the first few image orders, are shown in Fig.~\ref{fig:PhGeo}.
They provide the basis for understanding the relative winding of photon paths around the black hole.

\begin{figure*}[!htb]
  \centering
  \includegraphics[width=0.49\textwidth]{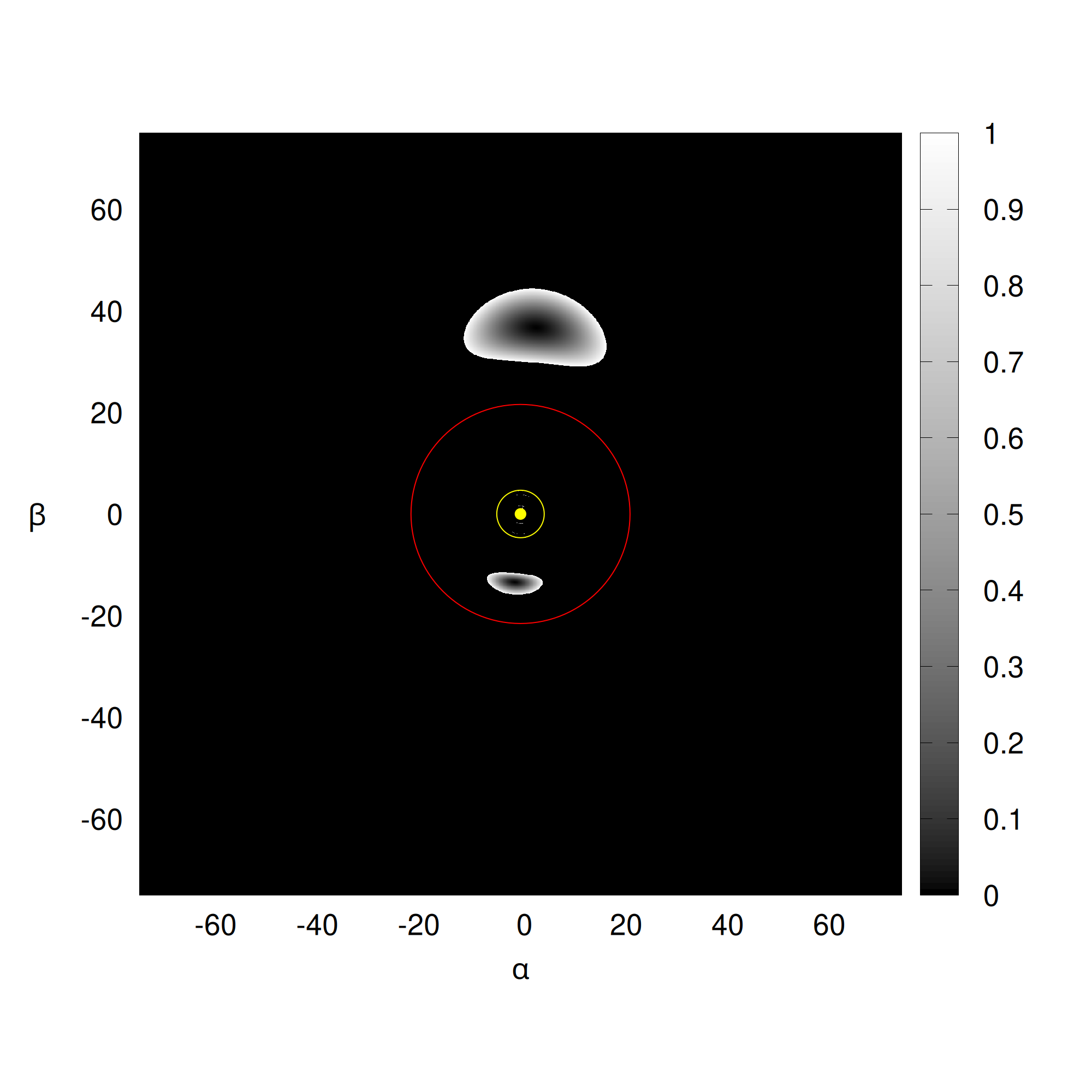}\hfill
  \includegraphics[width=0.49\textwidth]{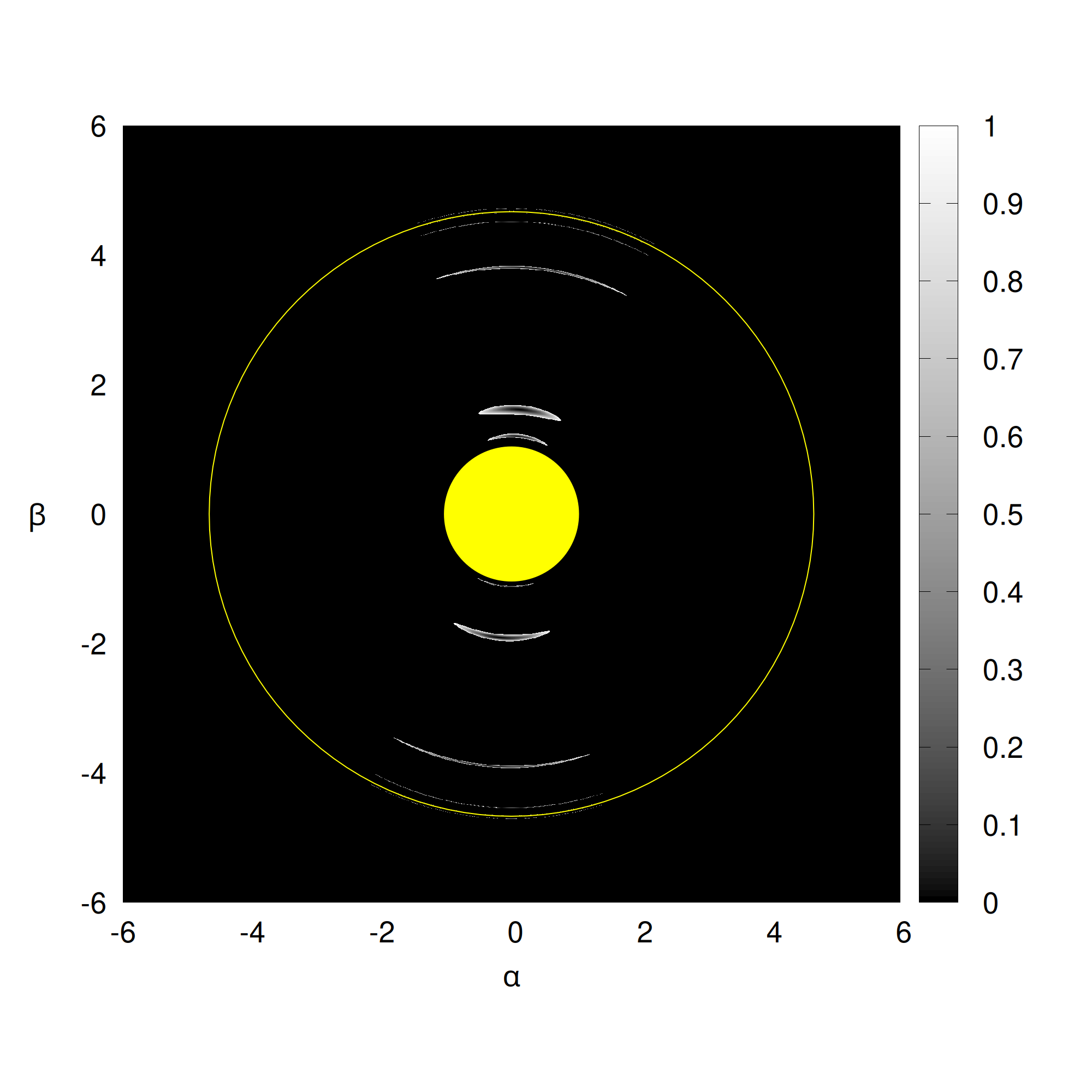}
  \caption{Ray-traced images of an extended circular source lensed by a Proca black hole
  with parameters $Q=-0.01$, $\alpha=-0.01$, and $\beta=1.02$.
  The grayscale encodes the radial distance from the center of the source, with white corresponding to its edge.
  Left panel: full image. Right panel: magnified central region of the left image. 
  Colored reference curves, obtained independently from ray tracing, are overplotted:
  the red ring marks the Einstein ring, 
  the yellow ring indicates the position of the outer photon orbit (present only in the Proca case), 
  and the yellow circle shows the black hole shadow, with its boundary corresponding to the inner photon orbit.
  Additional higher-order source images are visible, since the impact parameters corresponding to the source edges
  span a broader range than in the Schwarzschild case. With increasing order they approach the photon orbit values
  and eventually merge with the photon ring, becoming indistinguishable.
  Axes $(\alpha,\beta)$ are observer’s sky coordinates in units of $M$.}
  \label{fig:lens_proca}
\end{figure*}

The impact parameters, turning-point radii, swept angles, travel times, and relative time delays 
for different image orders, together with the color key used in Fig.~\ref{fig:PhGeo}, 
are summarized in Table~\ref{tab:time_del}.

\begin{table*}[!htb]
\centering
\caption{Color key and photon trajectories characteristics (see Fig.~\ref{fig:PhGeo}) for the 
source--observer configuration at $r_s=r_o=200M$. 
Times are given in units of $M$. 
The relative delay is defined as 
$\Delta t_{\rm rel}^{(n)} \equiv \Delta t^{(n)} - \Delta t^{(0)}$.}
\label{tab:time_del}
\begin{tabular}{l c c c c c c}
\hline\hline
Color / Style & Order $n$ & $l^{(n)}$ & $r_{\rm p}^{(n)}$ & $\Delta\phi^{(n)}$ & $\Delta t^{(n)}$ & $\Delta t_{\rm rel}^{(n)}$ \\
\hline
    Red, Solid        & 0 & 36.6725 & 35.6284 & $(\pi-0.25)$      & 405.431 &   0.0000 \\
    Blue, Solid       & 1 & 13.4975 & 12.3585 & $-(\pi+0.25)$     & 416.819 &  11.3877 \\
    Green, Solid      & 2 &  4.7206 &  2.6143 & $(3\pi-0.25)$     & 449.721 &  44.2903 \\
    Black, Solid      & 3 &  4.7089 &  2.5711 & $-(3\pi+0.25)$    & 452.078 &  46.6476 \\
    Pink, Solid       & 4 &  4.6801 &  2.3832 & $(5\pi-0.25)$     & 479.160 &  73.7287 \\
    Light Blue, Solid & 5 &  4.6800 &  2.3785 & $-(5\pi+0.25)$    & 481.493 &  76.0616 \\
    Green, Dotted     & 2 &  1.2045 &  0.3818 & $(3\pi-0.25)$     & 624.658 & 219.2270 \\
    Green, Dashed     & 2 &  1.6196 &  0.4726 & $(3\pi-0.25)$     & 624.715 & 219.2840 \\
    Black, Dotted     & 3 &  1.1209 &  0.3545 & $-(3\pi+0.25)$    & 625.236 & 219.8050 \\
    Black, Dashed     & 3 &  1.9160 &  0.5186 & $-(3\pi+0.25)$    & 625.603 & 220.1720 \\
    Pink, Dotted      & 4 &  1.0361 &  0.3039 & $(5\pi-0.25)$     & 631.314 & 225.8830 \\
    Light Blue, Dotted& 5 &  1.0359 &  0.3032 & $-(5\pi+0.25)$    & 631.832 & 226.4010 \\
    Pink, Dashed      & 4 &  3.8107 &  0.6849 & $(5\pi-0.25)$     & 642.938 & 237.5070 \\
    Light Blue, Dashed& 5 &  3.9076 &  0.6904 & $-(5\pi+0.25)$    & 644.868 & 239.4370 \\
\hline\hline
\end{tabular}
\end{table*}

\begin{figure*}[t]
  \centering
  \includegraphics[width=0.49\textwidth]{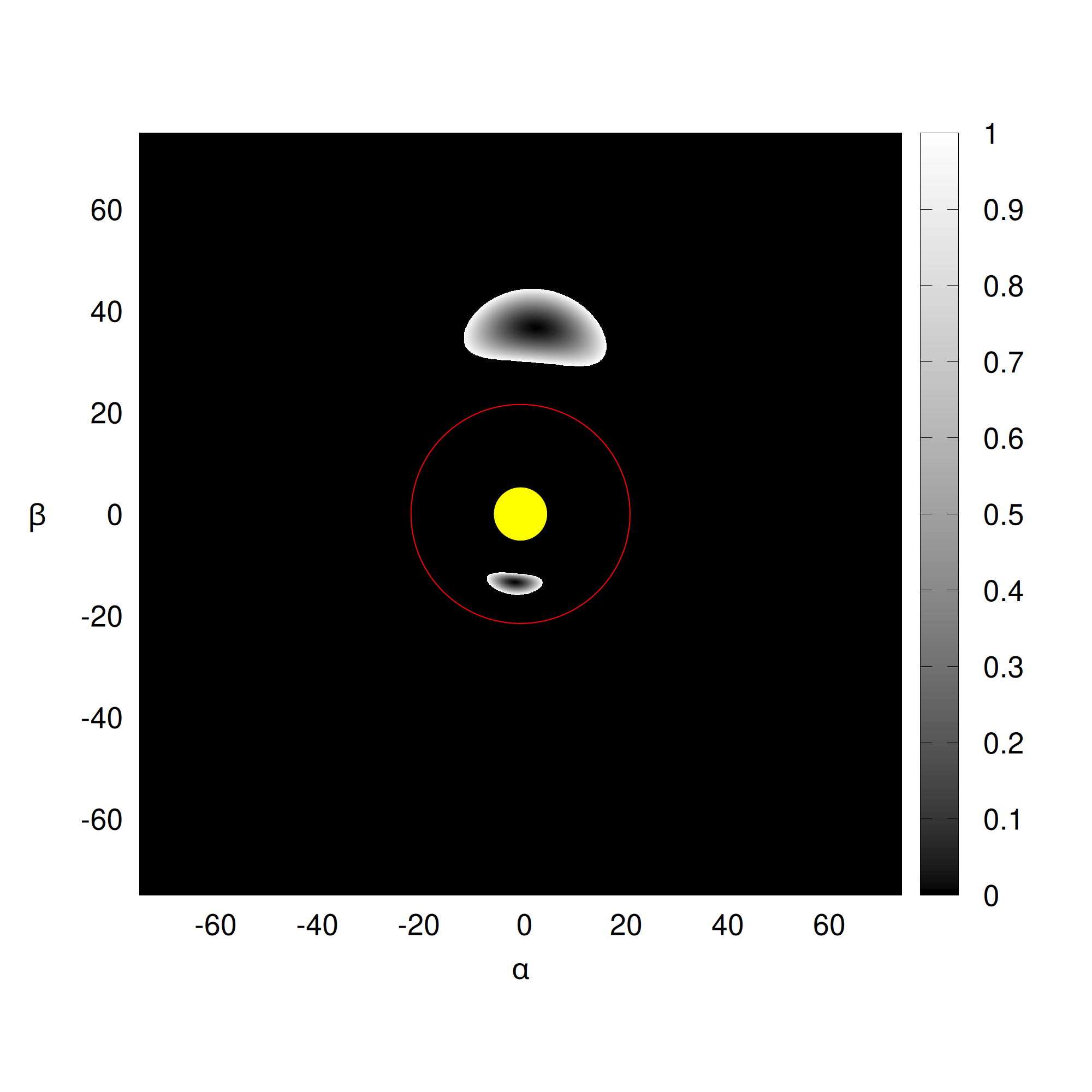}\hfill
  \includegraphics[width=0.49\textwidth]{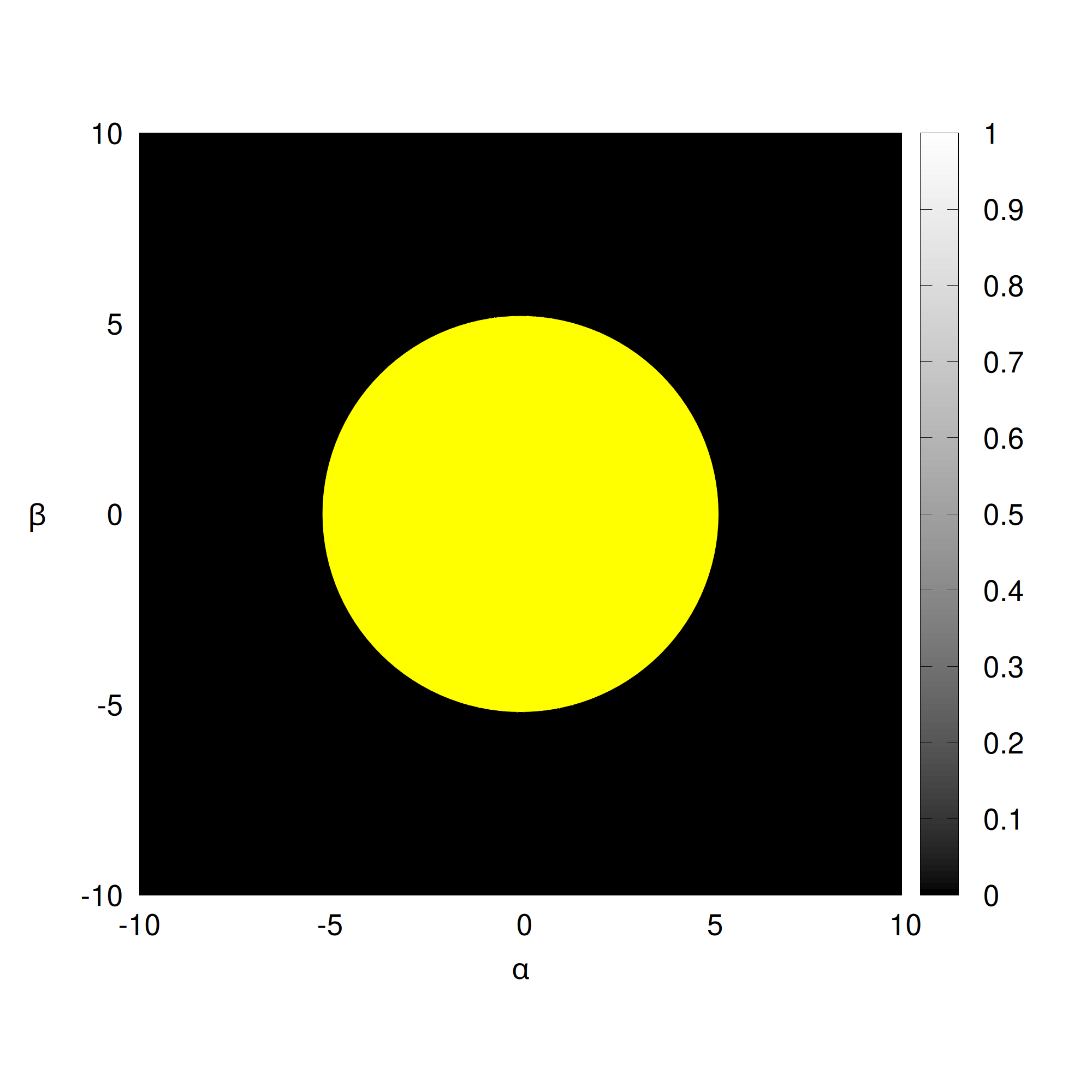}
  \caption{Same as Fig.~\ref{fig:lens_proca}, but for a Schwarzschild black hole. 
  Left panel: full image. Right panel: magnified central region of the left image. 
  Colored reference curves, obtained independently from ray tracing, are overplotted:
  the red ring marks the Einstein ring, and the yellow (white at the moment) circle indicates the black hole shadow,
  with its boundary corresponding to the photon orbit. For Schwarzschild, the impact parameters of higher-order images of the source differ only slightly from the critical photon orbit,
  which makes these images indistinguishable in the reconstructed picture.}
  \label{fig:lens_schw}
\end{figure*}

Using the ray-tracing method (details in Appendix~\ref{app:raytracing}), we reconstruct the images of the extended source as seen by a distant observer (Fig.~\ref{fig:lens_proca} for the Proca black hole and, for comparison, Fig.~\ref{fig:lens_schw} for the Schwarzschild black hole). The grayscale pattern encodes the radial distance from the center of the source, while the overplotted colored circles serve as references based on values obtained in previous sections rather than direct outputs of the ray-tracing procedure. Specifically, the red ring indicates the Einstein ring, the yellow ring (present only in the Proca case) marks the position of the outer photon orbit, and the yellow circle corresponds to the shadow boundary determined by the inner photon orbit of Proca black hole, and Single photon orbit in  Schwarzschild case.

In the Proca case, an additional distinctive feature is the visibility of higher-order source images: since the impact parameters associated with the source edges span a broader range than in the Schwarzschild spacetime, these images remain resolvable at lower orders. With increasing order they approach the photon orbit values and eventually merge with the photon ring, becoming indistinguishable. In the Schwarzschild case, by contrast, the impact parameters of higher-order images differ only slightly from the critical photon orbit already at low orders, which makes them indistinguishable in the reconstructed images. The visibility of higher-order images can therefore be regarded as one of the distinctive observable signatures of the Proca black hole relative to the Schwarzschild case.

Knowing the swept angle $\Delta\phi(l)$ as a function of the impact parameter $l$,
one can determine how the finite angular size of the source $\Delta\psi$ is mapped 
to the corresponding spread in the impact parameter, $\Delta l^{(n)}$,
for images of different orders. The results are summarized in Table~\ref{tab:phi_spread}.

\begin{table*}[!htb]
  \centering
  \caption{Mapping of the finite source size $\Delta\psi$
  into the spread of impact parameters $\Delta l^{(n)}$ for images of different orders~$n$.
  The spread in impact parameters is defined as 
  $\Delta l^{(n)} \equiv |l_{\max}^{(n)} - l_{\min}^{(n)}|$,
  while the source angular size is defined via
  $\Delta\psi \equiv |\Delta\phi(l_{\max}^{(n)}) - \Delta\phi(l_{\min}^{(n)})|$.}
  \begin{tabular}{c c c c c c c}
    \hline\hline
    Order $n$ & $\Delta\phi(l_{\min}^{(n)})$  & $\Delta\phi(l_{\max}^{(n)})$  & $\Delta\psi$  & $l_{\min}^{(n)}$  & $l_{\max}^{(n)}$  & $\Delta l^{(n)}$ \\
    \hline
     0 & $\pi-0.15$   & $\pi-0.35$   & 0.2 & 29.8148 & 44.3124 & 14.4976 \\
     1 & $\pi+0.15$   & $\pi+0.35$   & 0.2 & 15.9842 & 11.7032 &  4.2810 \\
     2 & $3\pi-0.15$  & $3\pi-0.35$  & 0.2 &  1.6884 &  1.5381 &  0.1503 \\
     3 & $3\pi+0.15$  & $3\pi+0.35$  & 0.2 &  1.8634 &  1.9667 &  0.1033 \\
     2 & $3\pi-0.15$  & $3\pi-0.35$  & 0.2 &  1.1788 &  1.2426 &  0.0638 \\
     5 & $5\pi-0.15$  & $5\pi-0.35$  & 0.2 &  3.8308 &  3.7903 &  0.0405 \\
     4 & $5\pi+0.15$  & $5\pi+0.35$  & 0.2 &  3.8889 &  3.9259 &  0.0370 \\
     3 & $3\pi+0.15$  & $3\pi+0.35$  & 0.2 &  1.1315 &  1.1118 &  0.0196 \\
     2 & $3\pi-0.15$  & $3\pi-0.35$  & 0.2 &  4.7179 &  4.7234 &  0.0055 \\
     3 & $3\pi+0.15$  & $3\pi+0.35$  & 0.2 &  4.7109 &  4.7070 &  0.0039 \\
     5 & $5\pi-0.15$  & $5\pi-0.35$  & 0.2 &  1.0361 &  1.0362 &  0.00010 \\
     5 & $5\pi-0.15$  & $5\pi-0.35$  & 0.2 &  4.6801 &  4.6802 &  0.00008 \\
     4 & $5\pi+0.15$  & $5\pi+0.35$  & 0.2 &  4.6800 &  4.6799 &  0.00007 \\
     4 & $5\pi+0.15$  & $5\pi+0.35$  & 0.2 &  1.0360 &  1.0359 &  0.00007 \\
    \hline\hline
  \end{tabular}
  \label{tab:phi_spread}
\end{table*}

\section{Conclusions}\label{sec:conclusions}

While key optical phenomena such as gravitational lensing, time delay, and, most importantly, black-hole shadows have been extensively studied across a wide range of gravitational theories and configurations with matter fields \cite{Ali:2025rlb,Konoplya:2019goy,Konoplya:2019xmn,Tsukamoto:2017fxq,Liu:2020ola,Konoplya:2024lch,Konoplya:2025mvj,Carballo-Rubio:2018jzw,Afrin:2021imp,Konoplya:2025nqv,Cunha:2016bjh,Younsi:2016azx,Konoplya:2019sns,Cano:2019ore,Wang:2017hjl,Held:2019xde,Younsi:2021dxe,Contreras:2019cmf,Bonanno:2025dry,Khodadi:2020gns}, including theories with higher-curvature corrections \cite{Konoplya:2020bxa,Konoplya:2019fpy,Hennigar:2018hza} (see also reviews \cite{Cunha:2018acu,Perlick:2021aok,Mizuno:2018lxz,Psaltis:2018xkc}), to the best of our knowledge no qualitatively new effects such as those associated with the emergence of a double-peak effective potential have been observed in these scenarios without introducing additional unusual matter components (for instance, local matter overdensities producing extra bumps in the potential for exotic matter supporting a wormhole). However, when the matter content is suitably chosen - potentially not involving even a mild violation of the energy conditions ~\cite{Guo:2022ghl} - a second peak may appear in the effective potential governing particle motion.  In the present work, we have demonstrated that the primary hair arising from the Proca field can itself generate a double-peak structure in the potential, thereby drastically altering the dynamics of both massless and massive particles, and consequently, all associated optical properties.

In this work, we have investigated the impact of primary Proca hair on the optical properties of black holes. Our analysis shows that the presence of the Proca field can drastically reshape the effective potential, generating a characteristic double-peak structure. This feature leads to the emergence of multiple circular orbits for both massless and massive particles and, consequently, to qualitatively different optical phenomena.

Our main findings can be summarized as follows:
\begin{itemize}
\item For a certain range of parameters, the black-hole model admits not only the standard photon sphere but also an additional, inner photon orbit, resulting in a richer orbital structure.
\item The coexistence of multiple photon spheres gives rise to modified lensing properties and shadows with a two-boundary structure, as well as additional inner rings that have no analogue in the Schwarzschild case or, to the best of our knowledge, in other black-hole solutions in alternative theories of gravity.
\item Time-delay effects and swept angles of photon trajectories display significant departures from standard black-hole expectations, reflecting the influence of the Proca-induced double-barrier potential.
\item Effective potentials for massive particles also acquire qualitatively new features, altering the stability regions for circular time-like geodesics.
\end{itemize}

Taken together, these results demonstrate that the presence of Proca hair imprints clear optical signatures, potentially observable in black-hole imaging and strong-field lensing data. Such effects could complement gravitational-wave probes in testing the existence of additional degrees of freedom beyond general relativity. A promising direction for future work is the extension of this analysis to rotating backgrounds.

\acknowledgments{The authors gratefully acknowledge the support of the Research Centre for Theoretical Physics and Astrophysics, 
Institute of Physics, Silesian University in Opava. 
D.O. further acknowledges financial support from the internal grant of the Silesian University in Opava, 
SGS/24/2024 Astrophysical processes in strong gravitational and electromagnetic fields of compact objects.}

\appendix
\section{Ray-tracing}
\label{app:raytracing}
To produce the lensing images discussed in Sec.~\ref{subsec:grav_lens}, 
we implemented a standard numerical ray-tracing scheme based on the geodesic equations 
presented in Sec.~\ref{sec:eq_motion}. The observer is placed at a fixed radial coordinate $r_o = 200M$,
with the image plane (screen) discretized in Cartesian coordinates $(\alpha,\beta)$,
each pixel corresponding to the initial direction of a photon launched backwards into the spacetime.
Photon trajectories are integrated using an adaptive step Runge--Kutta algorithm of order~5(4) \cite{galassi2002gnu},
and the integration is terminated once the photon either crosses the horizon, escapes to infinity (escapes the sphere $r_s$),
or intersects the source surface. The source is modeled as a finite circular patch of angular radius $\Delta\psi = 0.1$, located on a sphere of radius $r_s = 200M$. For visualization, the radial distance from the source center is encoded as a grayscale gradient, with white corresponding to the edge, in place of the more common representation by concentric rings.
For each trajectory we record the swept angle $\Delta\phi$, the travel time $\Delta t$, and, when present, the turning point radius $r_{\rm p}$.
These quantities allow us to determine the image order, the corresponding impact parameters,
and the mapping of the angular size of the source $\Delta\psi$ to the spread of the impact parameters corresponding to the edges of the source $\Delta l$, as summarized in Tables~\ref{tab:time_del} and~\ref{tab:phi_spread}.

\bibliography{bibliography}

\begin{thebibliography}{45}%
\makeatletter
\providecommand \@ifxundefined [1]{%
 \@ifx{#1\undefined}
}%
\providecommand \@ifnum [1]{%
 \ifnum #1\expandafter \@firstoftwo
 \else \expandafter \@secondoftwo
 \fi
}%
\providecommand \@ifx [1]{%
 \ifx #1\expandafter \@firstoftwo
 \else \expandafter \@secondoftwo
 \fi
}%
\providecommand \natexlab [1]{#1}%
\providecommand \enquote  [1]{``#1''}%
\providecommand \bibnamefont  [1]{#1}%
\providecommand \bibfnamefont [1]{#1}%
\providecommand \citenamefont [1]{#1}%
\providecommand \href@noop [0]{\@secondoftwo}%
\providecommand \href [0]{\begingroup \@sanitize@url \@href}%
\providecommand \@href[1]{\@@startlink{#1}\@@href}%
\providecommand \@@href[1]{\endgroup#1\@@endlink}%
\providecommand \@sanitize@url [0]{\catcode `\\12\catcode `\$12\catcode `\&12\catcode `\#12\catcode `\^12\catcode `\_12\catcode `\%12\relax}%
\providecommand \@@startlink[1]{}%
\providecommand \@@endlink[0]{}%
\providecommand \url  [0]{\begingroup\@sanitize@url \@url }%
\providecommand \@url [1]{\endgroup\@href {#1}{\urlprefix }}%
\providecommand \urlprefix  [0]{URL }%
\providecommand \Eprint [0]{\href }%
\providecommand \doibase [0]{http://dx.doi.org/}%
\providecommand \selectlanguage [0]{\@gobble}%
\providecommand \bibinfo  [0]{\@secondoftwo}%
\providecommand \bibfield  [0]{\@secondoftwo}%
\providecommand \translation [1]{[#1]}%
\providecommand \BibitemOpen [0]{}%
\providecommand \bibitemStop [0]{}%
\providecommand \bibitemNoStop [0]{.\EOS\space}%
\providecommand \EOS [0]{\spacefactor3000\relax}%
\providecommand \BibitemShut  [1]{\csname bibitem#1\endcsname}%
\let\auto@bib@innerbib\@empty
%</preamble>
\bibitem [{\citenamefont {Charmousis}\ \emph {et~al.}(2025)\citenamefont {Charmousis}, \citenamefont {Fernandes},\ and\ \citenamefont {Hassaine}}]{Charmousis:2025jpx}%
  \BibitemOpen
  \bibfield  {author} {\bibinfo {author} {\bibfnamefont {C.}~\bibnamefont {Charmousis}}, \bibinfo {author} {\bibfnamefont {P.~G.~S.}\ \bibnamefont {Fernandes}}, \ and\ \bibinfo {author} {\bibfnamefont {M.}~\bibnamefont {Hassaine}},\ }\href {\doibase 10.1103/9f2w-3kly} {\bibfield  {journal} {\bibinfo  {journal} {Phys. Rev. D}\ }\textbf {\bibinfo {volume} {111}},\ \bibinfo {pages} {124008} (\bibinfo {year} {2025})},\ \Eprint {http://arxiv.org/abs/2504.13084} {arXiv:2504.13084 [gr-qc]} \BibitemShut {NoStop}%
\bibitem [{\citenamefont {Konoplya}\ and\ \citenamefont {Zhidenko}(2025)}]{Konoplya:2025uiq}%
  \BibitemOpen
  \bibfield  {author} {\bibinfo {author} {\bibfnamefont {R.~A.}\ \bibnamefont {Konoplya}}\ and\ \bibinfo {author} {\bibfnamefont {A.}~\bibnamefont {Zhidenko}},\ }\href@noop {} {\  (\bibinfo {year} {2025})},\ \Eprint {http://arxiv.org/abs/2508.13069} {arXiv:2508.13069 [gr-qc]} \BibitemShut {NoStop}%
\bibitem [{\citenamefont {Akiyama}\ \emph {et~al.}(2019{\natexlab{a}})\citenamefont {Akiyama} \emph {et~al.}}]{EventHorizonTelescope:2019dse}%
  \BibitemOpen
  \bibfield  {author} {\bibinfo {author} {\bibfnamefont {K.}~\bibnamefont {Akiyama}} \emph {et~al.} (\bibinfo {collaboration} {Event Horizon Telescope}),\ }\href {\doibase 10.3847/2041-8213/ab0ec7} {\bibfield  {journal} {\bibinfo  {journal} {Astrophys. J. Lett.}\ }\textbf {\bibinfo {volume} {875}},\ \bibinfo {pages} {L1} (\bibinfo {year} {2019}{\natexlab{a}})},\ \Eprint {http://arxiv.org/abs/1906.11238} {arXiv:1906.11238 [astro-ph.GA]} \BibitemShut {NoStop}%
\bibitem [{\citenamefont {Akiyama}\ \emph {et~al.}(2022)\citenamefont {Akiyama} \emph {et~al.}}]{EventHorizonTelescope:2022wkp}%
  \BibitemOpen
  \bibfield  {author} {\bibinfo {author} {\bibfnamefont {K.}~\bibnamefont {Akiyama}} \emph {et~al.} (\bibinfo {collaboration} {Event Horizon Telescope}),\ }\href {\doibase 10.3847/2041-8213/ac6674} {\bibfield  {journal} {\bibinfo  {journal} {Astrophys. J. Lett.}\ }\textbf {\bibinfo {volume} {930}},\ \bibinfo {pages} {L12} (\bibinfo {year} {2022})},\ \Eprint {http://arxiv.org/abs/2311.08680} {arXiv:2311.08680 [astro-ph.HE]} \BibitemShut {NoStop}%
\bibitem [{\citenamefont {Akiyama}\ \emph {et~al.}(2019{\natexlab{b}})\citenamefont {Akiyama} \emph {et~al.}}]{EventHorizonTelescope:2019ggy}%
  \BibitemOpen
  \bibfield  {author} {\bibinfo {author} {\bibfnamefont {K.}~\bibnamefont {Akiyama}} \emph {et~al.} (\bibinfo {collaboration} {Event Horizon Telescope}),\ }\href {\doibase 10.3847/2041-8213/ab1141} {\bibfield  {journal} {\bibinfo  {journal} {Astrophys. J. Lett.}\ }\textbf {\bibinfo {volume} {875}},\ \bibinfo {pages} {L6} (\bibinfo {year} {2019}{\natexlab{b}})},\ \Eprint {http://arxiv.org/abs/1906.11243} {arXiv:1906.11243 [astro-ph.GA]} \BibitemShut {NoStop}%
\bibitem [{\citenamefont {Akiyama}\ \emph {et~al.}(2019{\natexlab{c}})\citenamefont {Akiyama} \emph {et~al.}}]{EventHorizonTelescope:2019pgp}%
  \BibitemOpen
  \bibfield  {author} {\bibinfo {author} {\bibfnamefont {K.}~\bibnamefont {Akiyama}} \emph {et~al.} (\bibinfo {collaboration} {Event Horizon Telescope}),\ }\href {\doibase 10.3847/2041-8213/ab0f43} {\bibfield  {journal} {\bibinfo  {journal} {Astrophys. J. Lett.}\ }\textbf {\bibinfo {volume} {875}},\ \bibinfo {pages} {L5} (\bibinfo {year} {2019}{\natexlab{c}})},\ \Eprint {http://arxiv.org/abs/1906.11242} {arXiv:1906.11242 [astro-ph.GA]} \BibitemShut {NoStop}%
\bibitem [{\citenamefont {Cunha}\ and\ \citenamefont {Herdeiro}(2018)}]{Cunha:2018acu}%
  \BibitemOpen
  \bibfield  {author} {\bibinfo {author} {\bibfnamefont {P.~V.~P.}\ \bibnamefont {Cunha}}\ and\ \bibinfo {author} {\bibfnamefont {C.~A.~R.}\ \bibnamefont {Herdeiro}},\ }\href {\doibase 10.1007/s10714-018-2361-9} {\bibfield  {journal} {\bibinfo  {journal} {Gen. Rel. Grav.}\ }\textbf {\bibinfo {volume} {50}},\ \bibinfo {pages} {42} (\bibinfo {year} {2018})},\ \Eprint {http://arxiv.org/abs/1801.00860} {arXiv:1801.00860 [gr-qc]} \BibitemShut {NoStop}%
\bibitem [{\citenamefont {Perlick}\ and\ \citenamefont {Tsupko}(2022)}]{Perlick:2021aok}%
  \BibitemOpen
  \bibfield  {author} {\bibinfo {author} {\bibfnamefont {V.}~\bibnamefont {Perlick}}\ and\ \bibinfo {author} {\bibfnamefont {O.~Y.}\ \bibnamefont {Tsupko}},\ }\href {\doibase 10.1016/j.physrep.2021.10.004} {\bibfield  {journal} {\bibinfo  {journal} {Phys. Rept.}\ }\textbf {\bibinfo {volume} {947}},\ \bibinfo {pages} {1} (\bibinfo {year} {2022})},\ \Eprint {http://arxiv.org/abs/2105.07101} {arXiv:2105.07101 [gr-qc]} \BibitemShut {NoStop}%
\bibitem [{\citenamefont {Akiyama}\ and\ \citenamefont {the Event Horizon Telescope~Collaboration}(2025)}]{Akiyama:2025aa55855}%
  \BibitemOpen
  \bibfield  {author} {\bibinfo {author} {\bibfnamefont {K.}~\bibnamefont {Akiyama}}\ and\ \bibinfo {author} {\bibnamefont {the Event Horizon Telescope~Collaboration}},\ }\href@noop {} {\bibfield  {journal} {\bibinfo  {journal} {Astronomy \& Astrophysics}\ }\textbf {\bibinfo {volume} {in press}} (\bibinfo {year} {2025})}\BibitemShut {NoStop}%
\bibitem [{\citenamefont {Lütfüoğlu}(2025)}]{Lutfuoglu:2025ldc}%
  \BibitemOpen
  \bibfield  {author} {\bibinfo {author} {\bibfnamefont {B.~C.}\ \bibnamefont {Lütfüoğlu}},\ }\href@noop {} {\bibfield  {journal} {\bibinfo  {journal} {International Journal of Gravitation and Theoretical Physics}\ }\textbf {\bibinfo {volume} {1}},\ \bibinfo {pages} {4} (\bibinfo {year} {2025})},\ \Eprint {http://arxiv.org/abs/2507.09246} {arXiv:2507.09246 [gr-qc]} \BibitemShut {NoStop}%
\bibitem [{\citenamefont {Fernandes}(2021)}]{Fernandes:2021dsb}%
  \BibitemOpen
  \bibfield  {author} {\bibinfo {author} {\bibfnamefont {P.~G.~S.}\ \bibnamefont {Fernandes}},\ }\href {\doibase 10.1103/PhysRevD.103.104065} {\bibfield  {journal} {\bibinfo  {journal} {Phys. Rev. D}\ }\textbf {\bibinfo {volume} {103}},\ \bibinfo {pages} {104065} (\bibinfo {year} {2021})},\ \Eprint {http://arxiv.org/abs/2105.04687} {arXiv:2105.04687 [gr-qc]} \BibitemShut {NoStop}%
\bibitem [{\citenamefont {Lu}\ and\ \citenamefont {Pang}(2020)}]{Lu:2020iav}%
  \BibitemOpen
  \bibfield  {author} {\bibinfo {author} {\bibfnamefont {H.}~\bibnamefont {Lu}}\ and\ \bibinfo {author} {\bibfnamefont {Y.}~\bibnamefont {Pang}},\ }\href {\doibase 10.1016/j.physletb.2020.135717} {\bibfield  {journal} {\bibinfo  {journal} {Phys. Lett. B}\ }\textbf {\bibinfo {volume} {809}},\ \bibinfo {pages} {135717} (\bibinfo {year} {2020})},\ \Eprint {http://arxiv.org/abs/2003.11552} {arXiv:2003.11552 [gr-qc]} \BibitemShut {NoStop}%
\bibitem [{\citenamefont {Kobayashi}(2020)}]{Kobayashi:2020wqy}%
  \BibitemOpen
  \bibfield  {author} {\bibinfo {author} {\bibfnamefont {T.}~\bibnamefont {Kobayashi}},\ }\href {\doibase 10.1088/1475-7516/2020/07/013} {\bibfield  {journal} {\bibinfo  {journal} {JCAP}\ }\textbf {\bibinfo {volume} {07}},\ \bibinfo {pages} {013} (\bibinfo {year} {2020})},\ \Eprint {http://arxiv.org/abs/2003.12771} {arXiv:2003.12771 [gr-qc]} \BibitemShut {NoStop}%
\bibitem [{\citenamefont {Fernandes}\ \emph {et~al.}(2020)\citenamefont {Fernandes}, \citenamefont {Carrilho}, \citenamefont {Clifton},\ and\ \citenamefont {Mulryne}}]{Fernandes:2020nbq}%
  \BibitemOpen
  \bibfield  {author} {\bibinfo {author} {\bibfnamefont {P.~G.~S.}\ \bibnamefont {Fernandes}}, \bibinfo {author} {\bibfnamefont {P.}~\bibnamefont {Carrilho}}, \bibinfo {author} {\bibfnamefont {T.}~\bibnamefont {Clifton}}, \ and\ \bibinfo {author} {\bibfnamefont {D.~J.}\ \bibnamefont {Mulryne}},\ }\href {\doibase 10.1103/PhysRevD.102.024025} {\bibfield  {journal} {\bibinfo  {journal} {Phys. Rev. D}\ }\textbf {\bibinfo {volume} {102}},\ \bibinfo {pages} {024025} (\bibinfo {year} {2020})},\ \Eprint {http://arxiv.org/abs/2004.08362} {arXiv:2004.08362 [gr-qc]} \BibitemShut {NoStop}%
\bibitem [{\citenamefont {Arnowitt}\ \emph {et~al.}(1960)\citenamefont {Arnowitt}, \citenamefont {Deser},\ and\ \citenamefont {Misner}}]{Arnowitt:1960es}%
  \BibitemOpen
  \bibfield  {author} {\bibinfo {author} {\bibfnamefont {R.~L.}\ \bibnamefont {Arnowitt}}, \bibinfo {author} {\bibfnamefont {S.}~\bibnamefont {Deser}}, \ and\ \bibinfo {author} {\bibfnamefont {C.~W.}\ \bibnamefont {Misner}},\ }\href {\doibase 10.1103/PhysRev.117.1595} {\bibfield  {journal} {\bibinfo  {journal} {Phys. Rev.}\ }\textbf {\bibinfo {volume} {117}},\ \bibinfo {pages} {1595} (\bibinfo {year} {1960})}\BibitemShut {NoStop}%
\bibitem [{\citenamefont {Stuchlík}\ \emph {et~al.}(2019)\citenamefont {Stuchlík}, \citenamefont {Schee},\ and\ \citenamefont {Ovchinnikov}}]{Stuchlik:2019apj}%
  \BibitemOpen
  \bibfield  {author} {\bibinfo {author} {\bibfnamefont {Z.}~\bibnamefont {Stuchlík}}, \bibinfo {author} {\bibfnamefont {J.}~\bibnamefont {Schee}}, \ and\ \bibinfo {author} {\bibfnamefont {D.}~\bibnamefont {Ovchinnikov}},\ }\href {\doibase 10.3847/1538-4357/ab55d5} {\bibfield  {journal} {\bibinfo  {journal} {The Astrophys. J.}\ }\textbf {\bibinfo {volume} {887}},\ \bibinfo {pages} {19} (\bibinfo {year} {2019})}\BibitemShut {NoStop}%
\bibitem [{\citenamefont {{Event Horizon Telescope Collaboration}}(2019)}]{EHT2019VI}%
  \BibitemOpen
  \bibfield  {author} {\bibinfo {author} {\bibnamefont {{Event Horizon Telescope Collaboration}}},\ }\href {\doibase 10.3847/2041-8213/ab1141} {\bibfield  {journal} {\bibinfo  {journal} {The Astrophysical Journal Letters}\ }\textbf {\bibinfo {volume} {875}},\ \bibinfo {pages} {L6} (\bibinfo {year} {2019})}\BibitemShut {NoStop}%
\bibitem [{\citenamefont {Akiyama}\ \emph {et~al.}(2024)\citenamefont {Akiyama}, \citenamefont {Alberdi}, \citenamefont {Alef},\ and\ \citenamefont {et~al.}}]{Akiyama2024_persistent}%
  \BibitemOpen
  \bibfield  {author} {\bibinfo {author} {\bibfnamefont {K.}~\bibnamefont {Akiyama}}, \bibinfo {author} {\bibfnamefont {A.}~\bibnamefont {Alberdi}}, \bibinfo {author} {\bibfnamefont {W.}~\bibnamefont {Alef}}, \ and\ \bibinfo {author} {\bibnamefont {et~al.}},\ }\href {\doibase 10.1051/0004-6361/202347932} {\bibfield  {journal} {\bibinfo  {journal} {Astronomy \& Astrophysics}\ }\textbf {\bibinfo {volume} {681}},\ \bibinfo {pages} {A79} (\bibinfo {year} {2024})}\BibitemShut {NoStop}%
\bibitem [{\citenamefont {Ali}\ \emph {et~al.}(2025)\citenamefont {Ali}, \citenamefont {Negi},\ and\ \citenamefont {Pant}}]{Ali:2025rlb}%
  \BibitemOpen
  \bibfield  {author} {\bibinfo {author} {\bibfnamefont {M.~S.}\ \bibnamefont {Ali}}, \bibinfo {author} {\bibfnamefont {A.}~\bibnamefont {Negi}}, \ and\ \bibinfo {author} {\bibfnamefont {S.}~\bibnamefont {Pant}},\ }\href@noop {} {\  (\bibinfo {year} {2025})},\ \Eprint {http://arxiv.org/abs/2509.03507} {arXiv:2509.03507 [gr-qc]} \BibitemShut {NoStop}%
\bibitem [{\citenamefont {Konoplya}\ and\ \citenamefont {Zhidenko}(2019)}]{Konoplya:2019goy}%
  \BibitemOpen
  \bibfield  {author} {\bibinfo {author} {\bibfnamefont {R.~A.}\ \bibnamefont {Konoplya}}\ and\ \bibinfo {author} {\bibfnamefont {A.}~\bibnamefont {Zhidenko}},\ }\href {\doibase 10.1103/PhysRevD.100.044015} {\bibfield  {journal} {\bibinfo  {journal} {Phys. Rev. D}\ }\textbf {\bibinfo {volume} {100}},\ \bibinfo {pages} {044015} (\bibinfo {year} {2019})},\ \Eprint {http://arxiv.org/abs/1907.05551} {arXiv:1907.05551 [gr-qc]} \BibitemShut {NoStop}%
\bibitem [{\citenamefont {Konoplya}(2020)}]{Konoplya:2019xmn}%
  \BibitemOpen
  \bibfield  {author} {\bibinfo {author} {\bibfnamefont {R.~A.}\ \bibnamefont {Konoplya}},\ }\href {\doibase 10.1016/j.physletb.2020.135363} {\bibfield  {journal} {\bibinfo  {journal} {Phys. Lett. B}\ }\textbf {\bibinfo {volume} {804}},\ \bibinfo {pages} {135363} (\bibinfo {year} {2020})},\ \Eprint {http://arxiv.org/abs/1912.10582} {arXiv:1912.10582 [gr-qc]} \BibitemShut {NoStop}%
\bibitem [{\citenamefont {Tsukamoto}(2018)}]{Tsukamoto:2017fxq}%
  \BibitemOpen
  \bibfield  {author} {\bibinfo {author} {\bibfnamefont {N.}~\bibnamefont {Tsukamoto}},\ }\href {\doibase 10.1103/PhysRevD.97.064021} {\bibfield  {journal} {\bibinfo  {journal} {Phys. Rev. D}\ }\textbf {\bibinfo {volume} {97}},\ \bibinfo {pages} {064021} (\bibinfo {year} {2018})},\ \Eprint {http://arxiv.org/abs/1708.07427} {arXiv:1708.07427 [gr-qc]} \BibitemShut {NoStop}%
\bibitem [{\citenamefont {Liu}\ \emph {et~al.}(2020)\citenamefont {Liu}, \citenamefont {Zhu}, \citenamefont {Wu}, \citenamefont {Jusufi}, \citenamefont {Jamil}, \citenamefont {Azreg-A{\"\i}nou},\ and\ \citenamefont {Wang}}]{Liu:2020ola}%
  \BibitemOpen
  \bibfield  {author} {\bibinfo {author} {\bibfnamefont {C.}~\bibnamefont {Liu}}, \bibinfo {author} {\bibfnamefont {T.}~\bibnamefont {Zhu}}, \bibinfo {author} {\bibfnamefont {Q.}~\bibnamefont {Wu}}, \bibinfo {author} {\bibfnamefont {K.}~\bibnamefont {Jusufi}}, \bibinfo {author} {\bibfnamefont {M.}~\bibnamefont {Jamil}}, \bibinfo {author} {\bibfnamefont {M.}~\bibnamefont {Azreg-A{\"\i}nou}}, \ and\ \bibinfo {author} {\bibfnamefont {A.}~\bibnamefont {Wang}},\ }\href {\doibase 10.1103/PhysRevD.101.084001} {\bibfield  {journal} {\bibinfo  {journal} {Phys. Rev. D}\ }\textbf {\bibinfo {volume} {101}},\ \bibinfo {pages} {084001} (\bibinfo {year} {2020})},\ \bibinfo {note} {[Erratum: Phys.Rev.D 103, 089902 (2021)]},\ \Eprint {http://arxiv.org/abs/2003.00477} {arXiv:2003.00477 [gr-qc]} \BibitemShut {NoStop}%
\bibitem [{\citenamefont {Konoplya}\ and\ \citenamefont {Stashko}(2025)}]{Konoplya:2024lch}%
  \BibitemOpen
  \bibfield  {author} {\bibinfo {author} {\bibfnamefont {R.~A.}\ \bibnamefont {Konoplya}}\ and\ \bibinfo {author} {\bibfnamefont {O.~S.}\ \bibnamefont {Stashko}},\ }\href {\doibase 10.1103/PhysRevD.111.104055} {\bibfield  {journal} {\bibinfo  {journal} {Phys. Rev. D}\ }\textbf {\bibinfo {volume} {111}},\ \bibinfo {pages} {104055} (\bibinfo {year} {2025})},\ \Eprint {http://arxiv.org/abs/2408.02578} {arXiv:2408.02578 [gr-qc]} \BibitemShut {NoStop}%
\bibitem [{\citenamefont {Konoplya}\ \emph {et~al.}(2025{\natexlab{a}})\citenamefont {Konoplya}, \citenamefont {Khrabustovskyi}, \citenamefont {K{\v{r}}{\'\i}{\v{z}}},\ and\ \citenamefont {Zhidenko}}]{Konoplya:2025mvj}%
  \BibitemOpen
  \bibfield  {author} {\bibinfo {author} {\bibfnamefont {R.~A.}\ \bibnamefont {Konoplya}}, \bibinfo {author} {\bibfnamefont {A.}~\bibnamefont {Khrabustovskyi}}, \bibinfo {author} {\bibfnamefont {J.}~\bibnamefont {K{\v{r}}{\'\i}{\v{z}}}}, \ and\ \bibinfo {author} {\bibfnamefont {A.}~\bibnamefont {Zhidenko}},\ }\href {\doibase 10.1088/1475-7516/2025/04/062} {\bibfield  {journal} {\bibinfo  {journal} {JCAP}\ }\textbf {\bibinfo {volume} {04}},\ \bibinfo {pages} {062} (\bibinfo {year} {2025}{\natexlab{a}})},\ \Eprint {http://arxiv.org/abs/2501.16134} {arXiv:2501.16134 [gr-qc]} \BibitemShut {NoStop}%
\bibitem [{\citenamefont {Carballo-Rubio}\ \emph {et~al.}(2018)\citenamefont {Carballo-Rubio}, \citenamefont {Di~Filippo}, \citenamefont {Liberati},\ and\ \citenamefont {Visser}}]{Carballo-Rubio:2018jzw}%
  \BibitemOpen
  \bibfield  {author} {\bibinfo {author} {\bibfnamefont {R.}~\bibnamefont {Carballo-Rubio}}, \bibinfo {author} {\bibfnamefont {F.}~\bibnamefont {Di~Filippo}}, \bibinfo {author} {\bibfnamefont {S.}~\bibnamefont {Liberati}}, \ and\ \bibinfo {author} {\bibfnamefont {M.}~\bibnamefont {Visser}},\ }\href {\doibase 10.1103/PhysRevD.98.124009} {\bibfield  {journal} {\bibinfo  {journal} {Phys. Rev. D}\ }\textbf {\bibinfo {volume} {98}},\ \bibinfo {pages} {124009} (\bibinfo {year} {2018})},\ \Eprint {http://arxiv.org/abs/1809.08238} {arXiv:1809.08238 [gr-qc]} \BibitemShut {NoStop}%
\bibitem [{\citenamefont {Afrin}\ \emph {et~al.}(2021)\citenamefont {Afrin}, \citenamefont {Kumar},\ and\ \citenamefont {Ghosh}}]{Afrin:2021imp}%
  \BibitemOpen
  \bibfield  {author} {\bibinfo {author} {\bibfnamefont {M.}~\bibnamefont {Afrin}}, \bibinfo {author} {\bibfnamefont {R.}~\bibnamefont {Kumar}}, \ and\ \bibinfo {author} {\bibfnamefont {S.~G.}\ \bibnamefont {Ghosh}},\ }\href {\doibase 10.1093/mnras/stab1260} {\bibfield  {journal} {\bibinfo  {journal} {Mon. Not. Roy. Astron. Soc.}\ }\textbf {\bibinfo {volume} {504}},\ \bibinfo {pages} {5927} (\bibinfo {year} {2021})},\ \Eprint {http://arxiv.org/abs/2103.11417} {arXiv:2103.11417 [gr-qc]} \BibitemShut {NoStop}%
\bibitem [{\citenamefont {Konoplya}\ \emph {et~al.}(2025{\natexlab{b}})\citenamefont {Konoplya}, \citenamefont {Stuchl{\'\i}k},\ and\ \citenamefont {Zhidenko}}]{Konoplya:2025nqv}%
  \BibitemOpen
  \bibfield  {author} {\bibinfo {author} {\bibfnamefont {R.~A.}\ \bibnamefont {Konoplya}}, \bibinfo {author} {\bibfnamefont {Z.}~\bibnamefont {Stuchl{\'\i}k}}, \ and\ \bibinfo {author} {\bibfnamefont {A.}~\bibnamefont {Zhidenko}},\ }\href@noop {} {\  (\bibinfo {year} {2025}{\natexlab{b}})},\ \Eprint {http://arxiv.org/abs/2509.03301} {arXiv:2509.03301 [gr-qc]} \BibitemShut {NoStop}%
\bibitem [{\citenamefont {Cunha}\ \emph {et~al.}(2016)\citenamefont {Cunha}, \citenamefont {Grover}, \citenamefont {Herdeiro}, \citenamefont {Radu}, \citenamefont {Runarsson},\ and\ \citenamefont {Wittig}}]{Cunha:2016bjh}%
  \BibitemOpen
  \bibfield  {author} {\bibinfo {author} {\bibfnamefont {P.~V.~P.}\ \bibnamefont {Cunha}}, \bibinfo {author} {\bibfnamefont {J.}~\bibnamefont {Grover}}, \bibinfo {author} {\bibfnamefont {C.}~\bibnamefont {Herdeiro}}, \bibinfo {author} {\bibfnamefont {E.}~\bibnamefont {Radu}}, \bibinfo {author} {\bibfnamefont {H.}~\bibnamefont {Runarsson}}, \ and\ \bibinfo {author} {\bibfnamefont {A.}~\bibnamefont {Wittig}},\ }\href {\doibase 10.1103/PhysRevD.94.104023} {\bibfield  {journal} {\bibinfo  {journal} {Phys. Rev. D}\ }\textbf {\bibinfo {volume} {94}},\ \bibinfo {pages} {104023} (\bibinfo {year} {2016})},\ \Eprint {http://arxiv.org/abs/1609.01340} {arXiv:1609.01340 [gr-qc]} \BibitemShut {NoStop}%
\bibitem [{\citenamefont {Younsi}\ \emph {et~al.}(2016)\citenamefont {Younsi}, \citenamefont {Zhidenko}, \citenamefont {Rezzolla}, \citenamefont {Konoplya},\ and\ \citenamefont {Mizuno}}]{Younsi:2016azx}%
  \BibitemOpen
  \bibfield  {author} {\bibinfo {author} {\bibfnamefont {Z.}~\bibnamefont {Younsi}}, \bibinfo {author} {\bibfnamefont {A.}~\bibnamefont {Zhidenko}}, \bibinfo {author} {\bibfnamefont {L.}~\bibnamefont {Rezzolla}}, \bibinfo {author} {\bibfnamefont {R.}~\bibnamefont {Konoplya}}, \ and\ \bibinfo {author} {\bibfnamefont {Y.}~\bibnamefont {Mizuno}},\ }\href {\doibase 10.1103/PhysRevD.94.084025} {\bibfield  {journal} {\bibinfo  {journal} {Phys. Rev. D}\ }\textbf {\bibinfo {volume} {94}},\ \bibinfo {pages} {084025} (\bibinfo {year} {2016})},\ \Eprint {http://arxiv.org/abs/1607.05767} {arXiv:1607.05767 [gr-qc]} \BibitemShut {NoStop}%
\bibitem [{\citenamefont {Konoplya}(2019)}]{Konoplya:2019sns}%
  \BibitemOpen
  \bibfield  {author} {\bibinfo {author} {\bibfnamefont {R.~A.}\ \bibnamefont {Konoplya}},\ }\href {\doibase 10.1016/j.physletb.2019.05.043} {\bibfield  {journal} {\bibinfo  {journal} {Phys. Lett. B}\ }\textbf {\bibinfo {volume} {795}},\ \bibinfo {pages} {1} (\bibinfo {year} {2019})},\ \Eprint {http://arxiv.org/abs/1905.00064} {arXiv:1905.00064 [gr-qc]} \BibitemShut {NoStop}%
\bibitem [{\citenamefont {Cano}\ and\ \citenamefont {Ruip{\'e}rez}(2019)}]{Cano:2019ore}%
  \BibitemOpen
  \bibfield  {author} {\bibinfo {author} {\bibfnamefont {P.~A.}\ \bibnamefont {Cano}}\ and\ \bibinfo {author} {\bibfnamefont {A.}~\bibnamefont {Ruip{\'e}rez}},\ }\href {\doibase 10.1007/JHEP05(2019)189} {\bibfield  {journal} {\bibinfo  {journal} {JHEP}\ }\textbf {\bibinfo {volume} {05}},\ \bibinfo {pages} {189} (\bibinfo {year} {2019})},\ \bibinfo {note} {[Erratum: JHEP 03, 187 (2020)]},\ \Eprint {http://arxiv.org/abs/1901.01315} {arXiv:1901.01315 [gr-qc]} \BibitemShut {NoStop}%
\bibitem [{\citenamefont {Wang}\ \emph {et~al.}(2017)\citenamefont {Wang}, \citenamefont {Chen},\ and\ \citenamefont {Jing}}]{Wang:2017hjl}%
  \BibitemOpen
  \bibfield  {author} {\bibinfo {author} {\bibfnamefont {M.}~\bibnamefont {Wang}}, \bibinfo {author} {\bibfnamefont {S.}~\bibnamefont {Chen}}, \ and\ \bibinfo {author} {\bibfnamefont {J.}~\bibnamefont {Jing}},\ }\href {\doibase 10.1088/1475-7516/2017/10/051} {\bibfield  {journal} {\bibinfo  {journal} {JCAP}\ }\textbf {\bibinfo {volume} {10}},\ \bibinfo {pages} {051} (\bibinfo {year} {2017})},\ \Eprint {http://arxiv.org/abs/1707.09451} {arXiv:1707.09451 [gr-qc]} \BibitemShut {NoStop}%
\bibitem [{\citenamefont {Held}\ \emph {et~al.}(2019)\citenamefont {Held}, \citenamefont {Gold},\ and\ \citenamefont {Eichhorn}}]{Held:2019xde}%
  \BibitemOpen
  \bibfield  {author} {\bibinfo {author} {\bibfnamefont {A.}~\bibnamefont {Held}}, \bibinfo {author} {\bibfnamefont {R.}~\bibnamefont {Gold}}, \ and\ \bibinfo {author} {\bibfnamefont {A.}~\bibnamefont {Eichhorn}},\ }\href {\doibase 10.1088/1475-7516/2019/06/029} {\bibfield  {journal} {\bibinfo  {journal} {JCAP}\ }\textbf {\bibinfo {volume} {06}},\ \bibinfo {pages} {029} (\bibinfo {year} {2019})},\ \Eprint {http://arxiv.org/abs/1904.07133} {arXiv:1904.07133 [gr-qc]} \BibitemShut {NoStop}%
\bibitem [{\citenamefont {Younsi}\ \emph {et~al.}(2023)\citenamefont {Younsi}, \citenamefont {Psaltis},\ and\ \citenamefont {{\"O}zel}}]{Younsi:2021dxe}%
  \BibitemOpen
  \bibfield  {author} {\bibinfo {author} {\bibfnamefont {Z.}~\bibnamefont {Younsi}}, \bibinfo {author} {\bibfnamefont {D.}~\bibnamefont {Psaltis}}, \ and\ \bibinfo {author} {\bibfnamefont {F.}~\bibnamefont {{\"O}zel}},\ }\href {\doibase 10.3847/1538-4357/aca58a} {\bibfield  {journal} {\bibinfo  {journal} {Astrophys. J.}\ }\textbf {\bibinfo {volume} {942}},\ \bibinfo {pages} {47} (\bibinfo {year} {2023})},\ \Eprint {http://arxiv.org/abs/2111.01752} {arXiv:2111.01752 [astro-ph.HE]} \BibitemShut {NoStop}%
\bibitem [{\citenamefont {Contreras}\ \emph {et~al.}(2020)\citenamefont {Contreras}, \citenamefont {Rinc{\'o}n}, \citenamefont {Panotopoulos}, \citenamefont {Bargue{\~n}o},\ and\ \citenamefont {Koch}}]{Contreras:2019cmf}%
  \BibitemOpen
  \bibfield  {author} {\bibinfo {author} {\bibfnamefont {E.}~\bibnamefont {Contreras}}, \bibinfo {author} {\bibfnamefont {{\'A}.}~\bibnamefont {Rinc{\'o}n}}, \bibinfo {author} {\bibfnamefont {G.}~\bibnamefont {Panotopoulos}}, \bibinfo {author} {\bibfnamefont {P.}~\bibnamefont {Bargue{\~n}o}}, \ and\ \bibinfo {author} {\bibfnamefont {B.}~\bibnamefont {Koch}},\ }\href {\doibase 10.1103/PhysRevD.101.064053} {\bibfield  {journal} {\bibinfo  {journal} {Phys. Rev. D}\ }\textbf {\bibinfo {volume} {101}},\ \bibinfo {pages} {064053} (\bibinfo {year} {2020})},\ \Eprint {http://arxiv.org/abs/1906.06990} {arXiv:1906.06990 [gr-qc]} \BibitemShut {NoStop}%
\bibitem [{\citenamefont {Bonanno}\ \emph {et~al.}(2025)\citenamefont {Bonanno}, \citenamefont {Konoplya}, \citenamefont {Oglialoro},\ and\ \citenamefont {Spina}}]{Bonanno:2025dry}%
  \BibitemOpen
  \bibfield  {author} {\bibinfo {author} {\bibfnamefont {A.}~\bibnamefont {Bonanno}}, \bibinfo {author} {\bibfnamefont {R.~A.}\ \bibnamefont {Konoplya}}, \bibinfo {author} {\bibfnamefont {G.}~\bibnamefont {Oglialoro}}, \ and\ \bibinfo {author} {\bibfnamefont {A.}~\bibnamefont {Spina}},\ }\href@noop {} {\  (\bibinfo {year} {2025})},\ \Eprint {http://arxiv.org/abs/2509.12469} {arXiv:2509.12469 [gr-qc]} \BibitemShut {NoStop}%
\bibitem [{\citenamefont {Khodadi}\ and\ \citenamefont {Saridakis}(2021)}]{Khodadi:2020gns}%
  \BibitemOpen
  \bibfield  {author} {\bibinfo {author} {\bibfnamefont {M.}~\bibnamefont {Khodadi}}\ and\ \bibinfo {author} {\bibfnamefont {E.~N.}\ \bibnamefont {Saridakis}},\ }\href {\doibase 10.1016/j.dark.2021.100835} {\bibfield  {journal} {\bibinfo  {journal} {Phys. Dark Univ.}\ }\textbf {\bibinfo {volume} {32}},\ \bibinfo {pages} {100835} (\bibinfo {year} {2021})},\ \Eprint {http://arxiv.org/abs/2012.05186} {arXiv:2012.05186 [gr-qc]} \BibitemShut {NoStop}%
\bibitem [{\citenamefont {Konoplya}\ and\ \citenamefont {Zinhailo}(2020)}]{Konoplya:2020bxa}%
  \BibitemOpen
  \bibfield  {author} {\bibinfo {author} {\bibfnamefont {R.~A.}\ \bibnamefont {Konoplya}}\ and\ \bibinfo {author} {\bibfnamefont {A.~F.}\ \bibnamefont {Zinhailo}},\ }\href {\doibase 10.1140/epjc/s10052-020-08639-8} {\bibfield  {journal} {\bibinfo  {journal} {Eur. Phys. J. C}\ }\textbf {\bibinfo {volume} {80}},\ \bibinfo {pages} {1049} (\bibinfo {year} {2020})},\ \Eprint {http://arxiv.org/abs/2003.01188} {arXiv:2003.01188 [gr-qc]} \BibitemShut {NoStop}%
\bibitem [{\citenamefont {Konoplya}\ \emph {et~al.}(2020)\citenamefont {Konoplya}, \citenamefont {Pappas},\ and\ \citenamefont {Zhidenko}}]{Konoplya:2019fpy}%
  \BibitemOpen
  \bibfield  {author} {\bibinfo {author} {\bibfnamefont {R.~A.}\ \bibnamefont {Konoplya}}, \bibinfo {author} {\bibfnamefont {T.}~\bibnamefont {Pappas}}, \ and\ \bibinfo {author} {\bibfnamefont {A.}~\bibnamefont {Zhidenko}},\ }\href {\doibase 10.1103/PhysRevD.101.044054} {\bibfield  {journal} {\bibinfo  {journal} {Phys. Rev. D}\ }\textbf {\bibinfo {volume} {101}},\ \bibinfo {pages} {044054} (\bibinfo {year} {2020})},\ \Eprint {http://arxiv.org/abs/1907.10112} {arXiv:1907.10112 [gr-qc]} \BibitemShut {NoStop}%
\bibitem [{\citenamefont {Hennigar}\ \emph {et~al.}(2018)\citenamefont {Hennigar}, \citenamefont {Poshteh},\ and\ \citenamefont {Mann}}]{Hennigar:2018hza}%
  \BibitemOpen
  \bibfield  {author} {\bibinfo {author} {\bibfnamefont {R.~A.}\ \bibnamefont {Hennigar}}, \bibinfo {author} {\bibfnamefont {M.~B.~J.}\ \bibnamefont {Poshteh}}, \ and\ \bibinfo {author} {\bibfnamefont {R.~B.}\ \bibnamefont {Mann}},\ }\href {\doibase 10.1103/PhysRevD.97.064041} {\bibfield  {journal} {\bibinfo  {journal} {Phys. Rev. D}\ }\textbf {\bibinfo {volume} {97}},\ \bibinfo {pages} {064041} (\bibinfo {year} {2018})},\ \Eprint {http://arxiv.org/abs/1801.03223} {arXiv:1801.03223 [gr-qc]} \BibitemShut {NoStop}%
\bibitem [{\citenamefont {Mizuno}\ \emph {et~al.}(2018)\citenamefont {Mizuno}, \citenamefont {Younsi}, \citenamefont {Fromm}, \citenamefont {Porth}, \citenamefont {De~Laurentis}, \citenamefont {Olivares}, \citenamefont {Falcke}, \citenamefont {Kramer},\ and\ \citenamefont {Rezzolla}}]{Mizuno:2018lxz}%
  \BibitemOpen
  \bibfield  {author} {\bibinfo {author} {\bibfnamefont {Y.}~\bibnamefont {Mizuno}}, \bibinfo {author} {\bibfnamefont {Z.}~\bibnamefont {Younsi}}, \bibinfo {author} {\bibfnamefont {C.~M.}\ \bibnamefont {Fromm}}, \bibinfo {author} {\bibfnamefont {O.}~\bibnamefont {Porth}}, \bibinfo {author} {\bibfnamefont {M.}~\bibnamefont {De~Laurentis}}, \bibinfo {author} {\bibfnamefont {H.}~\bibnamefont {Olivares}}, \bibinfo {author} {\bibfnamefont {H.}~\bibnamefont {Falcke}}, \bibinfo {author} {\bibfnamefont {M.}~\bibnamefont {Kramer}}, \ and\ \bibinfo {author} {\bibfnamefont {L.}~\bibnamefont {Rezzolla}},\ }\href {\doibase 10.1038/s41550-018-0449-5} {\bibfield  {journal} {\bibinfo  {journal} {Nature Astron.}\ }\textbf {\bibinfo {volume} {2}},\ \bibinfo {pages} {585} (\bibinfo {year} {2018})},\ \Eprint {http://arxiv.org/abs/1804.05812} {arXiv:1804.05812 [astro-ph.GA]} \BibitemShut {NoStop}%
\bibitem [{\citenamefont {Psaltis}(2019)}]{Psaltis:2018xkc}%
  \BibitemOpen
  \bibfield  {author} {\bibinfo {author} {\bibfnamefont {D.}~\bibnamefont {Psaltis}},\ }\href {\doibase 10.1007/s10714-019-2611-5} {\bibfield  {journal} {\bibinfo  {journal} {Gen. Rel. Grav.}\ }\textbf {\bibinfo {volume} {51}},\ \bibinfo {pages} {137} (\bibinfo {year} {2019})},\ \Eprint {http://arxiv.org/abs/1806.09740} {arXiv:1806.09740 [astro-ph.HE]} \BibitemShut {NoStop}%
\bibitem [{\citenamefont {Guo}\ \emph {et~al.}(2023)\citenamefont {Guo}, \citenamefont {Lu}, \citenamefont {Wang}, \citenamefont {Wu},\ and\ \citenamefont {Yang}}]{Guo:2022ghl}%
  \BibitemOpen
  \bibfield  {author} {\bibinfo {author} {\bibfnamefont {G.}~\bibnamefont {Guo}}, \bibinfo {author} {\bibfnamefont {Y.}~\bibnamefont {Lu}}, \bibinfo {author} {\bibfnamefont {P.}~\bibnamefont {Wang}}, \bibinfo {author} {\bibfnamefont {H.}~\bibnamefont {Wu}}, \ and\ \bibinfo {author} {\bibfnamefont {H.}~\bibnamefont {Yang}},\ }\href {\doibase 10.1103/PhysRevD.107.124037} {\bibfield  {journal} {\bibinfo  {journal} {Phys. Rev. D}\ }\textbf {\bibinfo {volume} {107}},\ \bibinfo {pages} {124037} (\bibinfo {year} {2023})},\ \Eprint {http://arxiv.org/abs/2212.12901} {arXiv:2212.12901 [gr-qc]} \BibitemShut {NoStop}%
\bibitem [{\citenamefont {Galassi}\ \emph {et~al.}(2002)\citenamefont {Galassi}, \citenamefont {Davies}, \citenamefont {Theiler}, \citenamefont {Gough}, \citenamefont {Jungman}, \citenamefont {Alken}, \citenamefont {Booth}, \citenamefont {Rossi},\ and\ \citenamefont {Ulerich}}]{galassi2002gnu}%
  \BibitemOpen
  \bibfield  {author} {\bibinfo {author} {\bibfnamefont {M.}~\bibnamefont {Galassi}}, \bibinfo {author} {\bibfnamefont {J.}~\bibnamefont {Davies}}, \bibinfo {author} {\bibfnamefont {J.}~\bibnamefont {Theiler}}, \bibinfo {author} {\bibfnamefont {B.}~\bibnamefont {Gough}}, \bibinfo {author} {\bibfnamefont {G.}~\bibnamefont {Jungman}}, \bibinfo {author} {\bibfnamefont {P.}~\bibnamefont {Alken}}, \bibinfo {author} {\bibfnamefont {M.}~\bibnamefont {Booth}}, \bibinfo {author} {\bibfnamefont {F.}~\bibnamefont {Rossi}}, \ and\ \bibinfo {author} {\bibfnamefont {R.}~\bibnamefont {Ulerich}},\ }\href@noop {} {\emph {\bibinfo {title} {GNU scientific library}}}\ (\bibinfo  {publisher} {Network Theory Limited Godalming},\ \bibinfo {year} {2002})\BibitemShut {NoStop}%
\end{thebibliography}%

\end{document}